\documentclass[11pt]{article}
\usepackage{putex}

\RequirePackage[a4paper,top=30.6mm,bottom=38.6mm,left=34.6mm,right=34.6mm,footskip=1.3cm]{geometry}
\usepackage{setspace}
\usepackage{booktabs}
\usepackage{cancel}
\usepackage{simplewick}
\usepackage{multirow}
\usepackage{comment}
\usepackage{bbold}
\usepackage{caption}
\usepackage{amsmath}
\usepackage{mathtools}
\usepackage{enumerate}
\usepackage{cite}
\usepackage{tensor}
\usepackage{slashed}
\usepackage[utf8]{inputenc}
\usepackage{rotating}
\usepackage{bigfoot}

\usepackage[
colorlinks=true,
linkcolor=black,
urlcolor=blue,
filecolor=black,
citecolor=red,
]{hyperref}

\usepackage[textsize=footnotesize]{todonotes}

\numberwithin{equation}{section}

\def \< {\left<}
\def \> {\right>}

\newcommand{\be}{\begin{equation}} \newcommand{\ee}{\end{equation}}
\newcommand{\bea}{\begin{eqnarray}}  \newcommand{\eea}{\end{eqnarray}}
\newcommand{\nn}{\nonumber}

\newcommand*{\del}{\partial}

\begin{document}

	\begin{center}        % Main title
		\Huge Goldstone Bosons and Convexity
	\end{center}
	
	\vspace{0.7cm}
	\begin{center}        % Authors
		{\large  Domenico Orlando$^{1,2}$ and Eran Palti$^3$}
	\end{center}
	
	\vspace{0.15cm}
	\begin{center}   
	\emph{${}^1$ INFN sezione di Torino.
via Pietro Giuria 1, 10125 Torino, Italy}\\[.3cm]     % Institutes
	\emph{${}^2$ Albert Einstein Center for Fundamental Physics
Institute for Theoretical Physics, University of Bern,
Sidlerstrasse 5, CH-3012 Bern, Switzerland}\\[.3cm]     % Institutes
		\emph{${}^3$ Department of Physics, Ben-Gurion University of the Negev, Be'er-Sheva 84105, Israel}\\[.3cm]
		\emph{}\\[.2cm]
		
		e-mails: \tt palti@bgu.ac.il, \;domenico.orlando@to.infn.it
	\end{center}
	
	\vspace{1cm}
	
	%%%%%%%%%%%%%%%%%%%%%%%%%%%%%%%%%%%%%%%%%%%%%%%
	%%%%%%%%%%%%%%%%%%%%%%%%%%%%%%%%%%%%%%%%%%%%%%%
	%%%%%%%%%%%%%%%%%%%%%%%%%%%%%%%%%%%%%%%%%%%%%%%
	%%%%%%%%%%%%%%%%%%%%%%%%%%%%%%%%%%%%%%%%%%%%%%%
	%%%%%%%%%%%%%%%%%%%%%%%%%%%%%%%%%%%%%%%%%%%%%%%
	%%%%%%%%%%%%%%%%%%%%%%%%%%%%%%%%%%%%%%%%%%%%%%%
	%%%%%%%%%%%%%%%%%%%%%%%%%%%%%%%%%%%%%%%%%%%%%%%
	%%%%%%%%%%%%%%%%%%%%%%%%%%%%%%%%%%%%%%%%%%%%%%%
	
	\begin{abstract}
	\noindent
	We study the spectrum of scalar charged operators in Conformal Field Theories (\textsc{cft}s) with a $U(1)$ global symmetry. The charged operators are dual, by the state-operator correspondence, to homogenous charged states on the sphere. Such states can break the $U(1)$ symmetry, and we define what we call the large $f$ regime in the \textsc{cft} as one where the  symmetry breaking scale is much higher than the scale of the \textsc{cft} sphere. In such a regime, there is (an approximate) Goldstone boson associated to the breaking. We show that consistency of the Goldstone boson physics implies that the spectrum of states, and therefore of operators, must be convex in charge. More precisely, we show that any family of operators of different charges, which are lowest dimension of their charge, and which additionally share the same realisation of the Goldstone boson in terms of the degrees of freedom of the \textsc{cft}, must be convex. 
	\end{abstract}
	
	\thispagestyle{empty}
	\clearpage
	
	\tableofcontents
	
	\setcounter{page}{1}
	
	%%%%%%%%%%%%%%%%%%%%%%%%%%%%%%%%%%%%%%%%%%%%%%%
	%%%%%%%%%%%%%%%%%%%%%%%%%%%%%%%%%%%%%%%%%%%%%%%
	%%%%%%%%%%%                 %%%%%%%%%%%%%%%%%%%
	%%%%%%%%%%%  DOCUMENT BODY  %%%%%%%%%%%%%%%%%%%
	%%%%%%%%%%%                 %%%%%%%%%%%%%%%%%%%
	%%%%%%%%%%%%%%%%%%%%%%%%%%%%%%%%%%%%%%%%%%%%%%%
	%%%%%%%%%%%%%%%%%%%%%%%%%%%%%%%%%%%%%%%%%%%%%%%
	%%%%%%%%%%%%%%%%%%%%%%%%%%%%%%%%%%%%%%%%%%%%%%%
	
	%\newpage

%%%%%%%%%%%%%%%%%%%%%%%%%%%%%%%%%%%%%%%%%%%%%%%
\section{Introduction}
\label{sec:intro}
%%%%%%%%%%%%%%%%%%%%%%%%%%%%%%%%%%%%%%%%%%%%%%%	
	
	In this paper we study the operator spectrum of Conformal Field Theories (\textsc{cft}s) which have a global $U(1)$ symmetry. Our primary tool is the state-operator correspondence, which maps the charged operators to charged states. The charged states (apart from some exceptional cases) break the $U(1)$ symmetry and so, in flat space, would have an associated Goldstone boson. The state-operator correspondence relies on considering the \textsc{cft} on a sphere though, and we can associate an energy scale to the sphere $\frac{1}{R}$. By continuity of physics, if the symmetry breaking scale $f$ is much larger than the sphere scale, we should recover the flat space physics and have a Goldstone boson. We therefore introduce what we call the large $f$ regime, where such a Goldstone boson is present in the theory. We then use the physics of the Goldstone boson to put constraints on the spectrum of states, and therefore of operators. In particular, we show that consistency of its physics implies a certain convexity of the spectrum of states/operators. 
	
	The large $f$ regime is a new type of regime where \textsc{cft}s can be studied. It is similar in spirit to the large charge regime which has been studied intensively in the past years, starting from the work in \cite{Hellerman:2015nra}. In particular, a large charge regime of the \textsc{cft} is also a large $f$ one, but not vice-versa: the large $f$ regime is less restrictive than the large charge one. In particular, the charge of the states/operators need not be larger than any other parameter in the theory, and there are large $f$ regimes even in perturbative regimes of \textsc{cft}s.  
	
	We show that, within a large $f$ regime of a \textsc{cft}, there are families of charged operators/states that are convex in charge. Such a family is defined as the lowest energy states, or lowest dimension operators, for a range of charges within the regime, which share the embedding of the Goldstone boson, or more precisely the symmetry breaking, into the degrees of freedom of the \textsc{cft}. In weakly-coupled \textsc{cft}s, such a family would be, for example, $\phi^n$, with $\phi$ being some charged field and $n$ a positive integer (within a certain range).\footnote{Such a family of operators is guaranteed to be also of smallest dimension only if $\phi$ is the field with the largest charge. Then convexity is guaranteed. Otherwise, convexity may or may not hold, depending on if there are operators of lower dimensions which have a different realisation of the Goldstone boson.}
	
	An important motivation for our work is the Charge Convexity Conjecture \cite{Aharony:2021mpc}. Let us label by $\Delta(Q)$ the dimension of the lowest-dimension operator of charge $Q$ under the $U(1)$ global symmetry in the \textsc{cft}, then the Charge Convexity Conjecture proposes~\cite{Aharony:2021mpc}
	\be
	\Delta\left( \left(m+n\right)\;q_0 \right) \geq \Delta\left( m\; q_0 \right) + \Delta\left( n\; q_0 \right) \;,
	\label{ccc}
	\ee
	where $m$ and $n$ are any positive integers, and $q_0$ is an integer of order one. More precisely, $q_0$ is conjectured to never be parametrically large. The Charge Convexity Conjecture has been studied further in~\cite{Dupuis:2021flq,Antipin:2021rsh,Moser:2021bes,Aalsma:2021qga,Watanabe:2022htq,Palti:2022unw}. Recently, a counter-example to the conjecture was found in \cite{Sharon:2023drx}. This counter-example is consistent with our results: the non-convexity arises from a setting where the lowest dimension charged operators are found in a parametrically large number of different families of operators, so which have different realisations of the Goldstone boson. Only at parametrically large charge does one find a family of operators which is of lowest dimension and shares a Goldstone boson realisation, that family must then be convex.
		
	 In this paper, we focus on the case where the operators in question are scalar, which means that the dual states are homogeneous. It is the scalar case which is the most interesting, since classically it is marginally convex. Of course, this does not imply that the \textsc{cft} should not have fermions, only that the lowest operators for a given charge transform as scalars (they may be scalar composites of fermions). 
	
	The Charge Convexity Conjecture is holographically related to the Weak Gravity Conjecture \cite{Arkani-Hamed:2006emk} in AdS. More precisely, it is related, though not identical, to the Positive Binding Conjecture in AdS \cite{Aharony:2021mpc}. This proposes that there should exist a charged particle which has positive self-binding energy.\footnote{See \cite{Andriolo:2022hax} for a calculation of such binding energies in AdS. Note, however, that there are important and relevant works on other proposed holographic dual formulations of the Weak Gravity Conjecture, see for example~\cite{Nakayama:2015hga,Montero:2016tif,Montero:2018fns}. Also, see \cite{Komargodski:2012ek} for earlier work discussing convexity at large spin, and the more recent results on large spin with global charge \cite{Cuomo:2022kio}.} 	It is the natural generalisation of the repulsive force version of the Weak Gravity Conjecture in flat space~\cite{Palti:2017elp,Heidenreich:2019zkl}. Convexity of charged operators in \textsc{cft}s is therefore an important property to understand in the context of the Swampland program \cite{Vafa:2005ui,Palti:2019pca,vanBeest:2021lhn,Palti:2020mwc}.

	The paper is structured as follows: In section~\ref{sec:ccp} we introduce the key ideas of symmetry breaking, the large $f$ regime, and Goldstone bosons. We then study the effective theory for the Goldstone boson and show that its consistency implies convexity of the action as a function of the chemical potential. In section \ref{sec:cco} we study how the convexity of the action in the chemical potential is mapped to convexity in charge of the states/operators. We discuss an example theory which manifests the general physics of this work in section \ref{sec:to2m}. In section \ref{sec:cce} we discuss an alternative, though similar in spirit, approach to convexity. Specifically, we show that convexity of states/operators is implied by requiring that the lowest energy state of a given charge {\it expectation value} is a charge eigenstate. We discuss our results in section \ref{sec:dis}.
	
%%%%%%%%%%%%%%%%%%%%%%%%%%%%%%%%%%%%%%%%%%%%%%%%%
\section{Convexity in the chemical potential}
\label{sec:ccp}
%%%%%%%%%%%%%%%%%%%%%%%%%%%%%%%%%%%%%%%%%%%%%%%%%

The state-operator correspondence allows us to study the operator dimension spectrum in terms of the energy spectrum of states. Classically, the Hamiltonian as a function of charge $Q$ is convex conjugate to, or the Legendre transform of, the Lagrangian as a function of the chemical potential $m$. In section \ref{sec:cco}, we expand on this relation. In this section, we study only the Lagrangian side in terms of the chemical potential. 

An important fact is that the state-operator correspondence requires that, in $d$ dimensions, the \textsc{cft} is placed on the cylinder \(\mathbb{R} \times S^{d-1}\). We take the sphere radius to be denoted by $R$. This introduces a scale into the theory, and all dimensionful quantities in the theory are measured relative to that scale. Generically, we work in units where $R=1$. However, when it is informative, we write the scale $R$ explicitly.

%%%%%%%%%%%%%%%%%%%%%%%%%%%%%%%%%%%%%%%%%%%%%%%%%
\subsection{Symmetry breaking and Goldstone bosons}
\label{sec:sbgb}
%%%%%%%%%%%%%%%%%%%%%%%%%%%%%%%%%%%%%%%%%%%%%%%%%

The \textsc{cft} has a global $SO(1,d+1)\times U(1)$ symmetry. The states we are considering are charged and so may break the $U(1)$ symmetry. We consider the case when the symmetry is broken, but in a spatially homogeneous way, so the state preserves the sphere symmetries. If the state is of minimal energy for its charge, then the $U(1)$ is not broken completely, but rather to a combination with the dilatation (time-translation) symmetry. This is called the superfluid breaking pattern \cite{Monin:2016jmo}, denoted as
\be
SO(1,d+1)\times U(1) \rightarrow SO(d) \times D' \;,
\ee 
where $D'$ is a combination of the original dilatation symmetry $D$, and the $U(1)$ symmetry. The specific combination is labelled by a parameter $m$, and this is what we call the chemical potential. In terms of operators, it means that the state is an eigenstate of the combination of operators $\hat{H} - m \hat{Q}$, with $\hat{H}$ being the Hamiltonian and $\hat{Q}$ the charge operator. On the Lagrangian side, it is a statement about the fields in the theory. Specifically, if the $U(1)$ acts on a field $\Pi$ non-linearly, so
\be
U(1) \;:\; \Pi \rightarrow \Pi + \xi \;,
\label{u1xi}
\ee
with $\xi$ being a constant, then in the Lagrangian the field may only appear through the combination 
\be
\chi \equiv m t + \Pi \;.
\label{chiPi}
\ee
Here $t$ denotes time, and $\chi$ is defined as this specific combination. It is informative to state the dimensions: here $\chi$, $\Pi$ are dimensionless, and $m$ has dimension one. 

%%%%%%%%%%%%%%%%%%%%%%%%%%%%%%%%%%%%%%%%%%%%%%%%%
\subsubsection{The large $f$ regime}
\label{sec:lfr}
%%%%%%%%%%%%%%%%%%%%%%%%%%%%%%%%%%%%%%%%%%%%%%%%%

Because we have a $U(1)$ symmetry, the constant $\xi$ in (\ref{u1xi}) must be periodic. This periodicity is associated to a dimensionful quantity $f$ which parameterises the scale of the symmetry breaking by the charged state. The field $\Pi$ should then be normalised to have unit periodicity by this scale. We therefore write (\ref{chiPi}) as 
\be
\chi = m t + \frac{\pi}{f} \;,
\ee
where now $f$ and $\pi$ have dimensions $\frac{d-2}{2}$. We refer to $\pi$ as the Goldstone boson, and work with it throughout the paper. In fact, since time $t$ cannot appear explicitly in the Lagrangian, the combination which appears is 
\be
Y_{\mu} = \partial_{\mu} \chi = m \delta_{\mu t} + \frac{\partial_{\mu}\pi}{f} \;,
\ee
from which one can construct relativistic invariants, such as
\bea
Y^2 \equiv Y^{\mu} Y_{\mu} &=& m^2 \left(1 + 2\frac{\dot{\pi}}{mf} + \frac{\partial_{\mu} \pi \partial^{\mu} \pi}{m^2f^2} \right)\;.
\label{defY2} 
\eea
Here, and henceforth, $\dot{\pi}$ denotes the time derivative $\partial_t \pi$.\footnote{Note that we work in signature $\eta_{\mu\nu} = \left[\;\mathrm{diag\;} \left(+1,-1,-1,...,-1\right) \;\right]_{\mu\nu}$.} 
We are interested in performing a general analysis of the Goldstone boson $\pi$ about the charged states. In order to do this we introduce {\it the large $f$ expansion}. We work in the regime
\be
\mathrm{Large}\;f\;\mathrm{:\;}\;\; f R^{\frac{d-2}{2}}  \gg 1 \;.
\label{largef}
\ee
The large $f$ regime is defined such that it allows for an expansion in powers of $\pi$. We can see this from (\ref{defY2}). Since $\pi$ is canonically normalised, its momentum modes on the sphere are quantised in units of $\frac{1}{R}$, which means order one in our units. We see then that the last two terms in (\ref{defY2}) are small.\footnote{We expect that $m$ cannot be parametrically small, since it is given by the derivative of the Hamiltonian with respect to charge, though have no proof. If there could exist a theory with $m$ parametrically small, then the large $f$ regime should be stated with the additional requirement $m f \gg 1$.} Note that this allows for an expansion in $\pi$, and $\pi$ only appears through its derivatives, but this is not always a derivative expansion in general. We may have terms that are higher order in derivatives, but with the same powers of $\pi$, that are important (for example, arising from terms like $\partial_{\mu}Y^{\mu}$).

It may be informative to have an example in mind for how the large $f$ regime can be realised. In section \ref{sec:to2m} we study such an example in detail: the $O(2)$-model. The theory is weakly-coupled, with a coupling $g$. Let us denote the charge of the state about which we are working as $Q$. There are two regimes of the theory of interest: the small charge regime $g Q \ll 1$ and the large charge regime $gQ \gg 1$. The former regime can be studied using perturbation theory, and the latter using the large charge expansion \cite{Hellerman:2015nra}. In table \ref{tab:fmQso2} we show how the parameters $m$ and $f$ behave in these two regimes. We see that both of the regimes are at large $f$. The large charge regime is automatically a large $f$ one, while the small charge regime requires an additional condition $Q \gg 1$. So the large $f$ regime can be thought of as a medium charge regime in this model. 
\begin{table}
\center
\def\arraystretch{2}
\begin{tabular}{|c|c|c|}
\hline
 & Small charge $gQ \ll 1$ & Large charge $gQ \gg 1$ \\
\hline
$f R^{\frac{d-2}{2}}$ & $\sqrt{Q}$ & $\frac{1}{\sqrt{g}}\left(gQ\right)^{\frac13} $ \\
\hline	
$m R$ & $1 + {\cal O}(gQ)$ & $\left(gQ\right)^{\frac13}$ \\
\hline
\end{tabular}
\caption{Table showing the behaviour of $f$ and $m$ with $g$ and $Q$ in the two charge regimes of the $O(2)$-model in $d=4-\epsilon$. The results are derived in section \ref{sec:to2m}. The large $f$ regime requires $Q \gg 1$ in the small charge regime, while for the large charge regime it is automatically satisfied because $g \ll 1$.}
\label{tab:fmQso2}	
\end{table}

%%%%%%%%%%%%%%%%%%%%%%%%%%%%%%%%%%%%%%%%%%%%%%%%%
\subsubsection{Goldstone bosons on a compact space}
\label{sec:ssb}
%%%%%%%%%%%%%%%%%%%%%%%%%%%%%%%%%%%%%%%%%%%%%%%%%

We are considering charged states, dual to the charged operators. The natural expectation is then that they break the global $U(1)$ symmetry spontaneously, and by Goldstone's theorem, should have a Goldstone boson which we can identify with $\pi$. This is indeed a correct expectation generically, but there are some requirements for this to hold. 

The primary cause for divergences from the generic expectation is that the \textsc{cft} is on a compact space. In is often stated, correctly, that on a compact space there is no Spontaneous Symmetry Breaking (\textsc{ssb}). Since \textsc{ssb} is a requirement for Goldstone's theorem, it is not clear that one expects a Goldstone boson. On the other hand, the symmetry breaking is described by the scale $f$, and the only other relevant dimensionful scale is $R$, and so we expect that in the large $f$ regime we should recover the flat space results and find a Goldstone boson. 

More precisely, when a global symmetry is broken spontaneously, there are a collection of states parametrised by the $U(1)$ symmetry. In flat space, this set of states are degenerate in energy. Moreover, each state is part of a separate Hilbert space, and moving between them is done by the Goldstone boson expectation value. On a compact space, the degeneracy can be lifted by global effects. The states now live in a single Hilbert space, and have non-vanishing wavefunction overlaps. So there is a non-vanishing probability of each such state decaying to the lowest energy one. The lowest energy state is the sum over all the states, and so does not break the $U(1)$ symmetry. Therefore, there is no \textsc{ssb} on a compact space. 

The wavefunction overlaps, or the global effects, vanish in the infinite volume limit. More precisely, they are exponentially suppressed by $f^2 R^{d-2}$. In the large $f$ regime (\ref{largef}), we can therefore neglect them. What this means is that we can study a state which does break the $U(1)$ symmetry, even if it is not the lowest energy one. The physics difference is only exponentially suppressed in a parameter we are sending to infinity. This is a very good approximation. 

There are some sporadic cases where there may not be a Goldstone boson, even in the large $f$ regime. This is true for any theory in two dimensions $d=2$. Another example set are theories of free fermions (see \cite{Komargodski:2021zzy}). However, an arbitrarily weak attractive interaction between the fermions will lead to a Cooper pair \cite{Cooper:1956zz}.

%%%%%%%%%%%%%%%%%%%%%%%%%%%%%%%%%%%%%%%%%%%%%%%%%
\subsection{The effective theory for the Goldstone boson }
\label{sec:effgb}
%%%%%%%%%%%%%%%%%%%%%%%%%%%%%%%%%%%%%%%%%%%%%%%%%

We are interested in the Lagrangian dependence on the chemical potential $m$. It is crucial to capture the full dependence on $m$, which means that if there are any fields which are sensitive to $m$, they must be integrated out. This way their implicit dependence on $m$ is manifested explicitly in the Lagrangian. We therefore are interested in an effective Lagrangian in which all the dependence on $m$ is explicit, and we denote this ${\cal L}_{\mathrm{eff}}(m)$. 

The Goldstone boson does not need to be integrated out in ${\cal L}_{\mathrm{eff}}(m)$. So in terms of the explicit dependence on $m$, we can write 
\be
{\cal L}_{\mathrm{eff}}\left(m\right)  = \left.{\cal L}_{\mathrm{eff}}\left(m,\pi\right) \right|_{\pi=0} \;.
\label{leffmpieq0}
\ee
The effective theory defined by integrating out all the fields apart from the Goldstone boson, so with the associated Lagrangian ${\cal L}_{\mathrm{eff}}\left(m,\pi\right)$, is the primary tool in this work. Specifically, we will use constraints on how this theory should behave to extract constraints on how $m$ must appear in ${\cal L}_{\mathrm{eff}}\left(m,\pi\right)$, and therefore also in ${\cal L}_{\mathrm{eff}}\left(m\right)$.

We are interested in writing an effective theory for a Goldstone field, and so the first thing we should consider is in what sense we have a Goldstone {\it field}. Goldstone's theorem implies the presence of soft momentum modes without a gap in the non-compact limit. Strictly speaking, it only guarantees a zero mode. For a field to exist, we may require a collection of momentum modes. Note that, in the compact case, the momentum modes are quantised in $\frac{1}{R}$. Of course, we are working with an effective theory about the charged state, and this theory is not expected to survive into the ultraviolet, so we do not expect or demand an infinite number of momentum modes in the field. We will consider two situations, studied in sections \ref{sec:gbg} and \ref{sec:gbl}. The first is when there is a sufficiently large gap between the energy of the first momentum mode of the Goldstone field, and the next charged state, or next massive field in the theory. In such a setting we can write an effective theory for the low momentum modes of $\pi$ which is an expansion in derivatives. We would then say that $\pi$ is a Goldstone field. The second case is when there is no gap between the first momentum mode and the next heaviest field. That case is more subtle, but we will show that even then one can write an effective theory for the Goldstone modes, say the zero mode, which will capture the physics we are after. 

There are cases when there is no gap at all between the Goldstone, even the zero mode, and other fields. The most common being supersymmetric theories where the charged operators of interest are BPS and have a moduli space. In such cases, we expect the analysis in this work to still hold, but have not studied it in depth. One reason is that BPS operators, charged under a single $U(1)$, will always have an exactly marginally-convex spectrum since their dimension grows exactly linearly with their charge.

%%%%%%%%%%%%%%%%%%%%%%%%%%%%%%%%%%%%%%%%%%%%%%%%%
\subsubsection{Goldstone boson with a gap}
\label{sec:gbg}
%%%%%%%%%%%%%%%%%%%%%%%%%%%%%%%%%%%%%%%%%%%%%%%%%

We consider here the setting when there is a gap between some non-trivial momentum modes for the Goldstone boson $\pi$, and the next heaviest field. In that case, in the large $f$ regime, we have a derivative expansion of ${\cal L}_{\mathrm{eff}}\left(m,\pi\right)$. We can expand as
\bea
{\cal L}_{\mathrm{eff}}\left(m,\pi\right) &=& \left.{\cal L}_{\mathrm{eff}}\left(m,\pi\right)\right|_{\pi=0} + \left.\frac{\partial {\cal L}_{\mathrm{eff}}\left(m,\pi\right)}{\partial \left(\partial_{\mu} \pi \right)} \right|_{\pi=0}\left(\partial_{\mu} \pi \right) \nn \\
& &+ \left.\frac12 \frac{\partial^2 {\cal L}_{\mathrm{eff}}\left(m,\pi\right)}{\partial \left(\partial_{\mu} \pi \right)\partial \left(\partial_{\nu} \pi \right)}\right|_{\pi=0}\left(\partial_{\mu} \pi \right) \left(\partial_{\nu} \pi \right) + ... \;.
\label{refhg}
\eea
The second term in (\ref{refhg}) vanishes due to the equations of motion for $\pi$. Recalling that ${\cal L}_{\mathrm{eff}}\left(m,\pi\right)$ is only a function of the combination $Y^2$ in (\ref{defY2}), we can write the differential operators acting on it as
\bea
\left.\frac12 \frac{\partial^2 {\cal L}_{\mathrm{eff}}\left(m,\pi\right)}{\partial \left(\partial_{\mu} \pi \right)\partial \left(\partial_{\nu} \pi \right)}\right|_{\pi=0} = \frac12 \frac{1}{f^2}\Bigg[\frac{g^{\mu\nu} }{m} \frac{\partial {\cal L}_{\mathrm{eff}}\left(m\right)}{\partial m} + \delta^{\mu0} \delta^{\nu0}  \left(\frac{\partial^2 {\cal L}_{\mathrm{eff}}\left(m\right)}{\partial m^2} - \frac{1}{m} \frac{\partial {\cal L}_{\mathrm{eff}}\left(m\right)}{\partial m}\right) \Bigg] \;.
\label{2dexY}
\eea
Where we have written the metric on $\mathbb{R} \times S^{d-1}$ as $g_{\mu\nu}$. We also decompose the indices $\mu = 0,i$, and note that $g_{00}=1$ and $g_{0i}=0$. Using (\ref{2dexY}) in (\ref{refhg}) then yields
\be
{\cal L}_{\mathrm{eff}}\left(m,\pi\right) = {\cal L}_{\mathrm{eff}}\left(m\right) + \frac{1}{2} \frac{1}{f^2}\Bigg[\frac{\partial^2 {\cal L}_{\mathrm{eff}}\left(m\right)}{\partial m^2} \dot{\pi}^2 - \frac{Q}{m}   \left(\partial_i \pi \right)\left(\partial_j \pi \right)g^{ij} \Bigg] \;,
\label{ltodi}
\ee
Here we have utilised a relation
\be
Q = \frac{\partial {\cal L}_{\mathrm{eff}}\left(m\right)}{\partial m} \;.
\ee 
This relation arises from the fact that the charge is the (Legendre) dual coordinate to the chemical potential. We discuss this in section \ref{sec:cco}. 

The terms in (\ref{ltodi}) are the leading derivative terms for the Goldstone field. The important point is that the coefficient of the time derivative is given by $\frac{\partial^2 {\cal L}_{\mathrm{eff}}\left(m\right)}{\partial m^2}$. For the theory to be well-behaved, we need this coefficient to be positive. We therefore recover
\be
\frac{\partial^2 {\cal L}_{\mathrm{eff}}\left(m\right)}{\partial m^2} > 0 \;.
\label{convexlagm}
\ee
Positivity of the second derivative, for a differentiable function, is equivalent to convexity. We therefore recover the result that the effective Lagrangian, in which the dependence on the chemical potential $m$ is manifested fully explicitly, must be convex in the chemical potential.\footnote{Note that because the state is homogeneous we can exchange convexity in the Lagrangian density ${\cal L}$ for convexity in the Lagrangian $L$.}

It is worth discussing in more detail why we would expect the coefficient in front of the time derivative to be positive. This is quite standard, theories with negative kinetic terms are notoriously difficult to make sense of. Nonetheless, this is an effective theory, and it could be that such seemingly pathological behaviour may be permitted within the momentum range where the theory holds. This seems to us very unlikely. However, it would be nice to prove a sharp contradiction which can be seen in the deep infrared. One way to do this is to consider the dispersion relation for the Goldstone modes. An expansion in eigenfunctions of the laplacian on the cylinder, which are the product of a plane wave and a hyperspherical harmonic
\begin{equation}
  \pi \sim e^{i \omega t} Y_{\ell m}(\Omega) \, ,  
\end{equation}
gives the dispersion relation
\be
\omega^2 = \left(\frac{\partial^2{\cal L}_{\mathrm{eff}}\left(m\right)}{\partial m^2}\right)^{-1} \frac{Q}{m V_{d-1}} \frac{\ell (\ell + d - 2)}{R^2}\; ,
\ee
where $V_{d-1}$ is the volume of the \(d-1\) sphere.
Since $Q>0$ and $m>0$, if the spectrum is not convex, $\frac{\partial^2{\cal L}_{\mathrm{eff}}\left(m\right)}{\partial m^2} < 0$, then $\omega$ obtains an imaginary component. An imaginary component in the dispersion relation implies an instability of the state, it is decaying in time. This is not consistent if the state is the minimal energy state within a given superselection (charge) sector.\footnote{Due to global instanton effects, on a compact space, the state which supports \textsc{ssb} is not of the lowest energy, but rather the lowest energy one is the sum over all such states. But this decay is {\it exponentially} small in $f$, and so is not what is being manifested here.} We therefore conclude that the Lagrangian density must be convex in $m$ {\it at least} around any minimal energy charged state. 

Finally, we note that we actually obtain another constraint from requiring that the speed of sound associated to the dispersion relation is subluminal. This gives
\be
\frac{\partial^2 L_{\mathrm{eff}}\left(m\right)}{\partial m^2} \geq \frac{Q}{m} \;.
\label{sublum}
\ee
We therefore find not only convexity, but one which grows with charge. 

%%%%%%%%%%%%%%%%%%%%%%%%%%%%%%%%%%%%%%%%%%%%%%%%%
\subsubsection{Goldstone boson with light states}
\label{sec:gbl}
%%%%%%%%%%%%%%%%%%%%%%%%%%%%%%%%%%%%%%%%%%%%%%%%%

The two-derivative analysis performed in section \ref{sec:gbg} holds as long as there is a mass gap between the lowest momentum modes of the Goldstone boson and the lightest massive fields in the theory. In this section we consider the case when there is no such gap. 

Let us recall that the Goldstone mode appears only through the combination $\partial_{\mu} \left( m t + \frac{\pi}{f} \right)$, and that we are working in the $f \gg 1$ regime. We are therefore performing an expansion in $\frac{\partial \pi}{f}$. Since the field $\pi$ is canonically normalised, its momentum modes are quantised in units of the (inverse) sphere radius $\frac{1}{R}$. There are some possibilities where such as expansion is not valid. A necessary condition is that there are additional fields that are as light as the radius scale $\frac{1}{R}$, in which case we do not have a derivative expansion even for the smallest non-trivial momentum mode of the Goldstone boson. Still, this is not necessarily a problem because those light fields may couple weakly to the Goldstone boson. So even if light, they would lead to suppressed corrections in the effective action. Therefore, for a derivative expansion to fail we require fields of mass of order $\frac{1}{R}$, whose coupling to the Goldstone boson is not suppressed by $f$. 

The starting point for the effective Lagrangian is considering integrating out all the massive fields, leaving an effective theory which depends only on $\pi$ and any additional light fields which couple to it. We denote these fields as $r^i$, and denote the effective Lagrangian, which depends on them, as ${\cal L}_{\pi,r}\left(m,\pi,r^i\right)$. We take these fields to be weakly-coupled, in that there is a well-defined field, even if it is an effective one.\footnote{It is not clear to us how to couple the Goldstone in general to a strongly-coupled sector. Note, however, that the analysis of the large $f$ limit, around (\ref{getnga}), implies that the Goldstone boson should couple to some combination of the degrees of freedom which has dimension very close to that of a weakly-coupled scalar, suggesting that this is the general situation.} We also take the $r^i$ as fields describing fluctuations about their expectation value, so that we can take the action quadratic in them at leading order. The effective Lagrangian admits an expansion in derivatives, since the $r^i$ are manifested explicitly and all other fields are massive. 

Before writing the expression for ${\cal L}_{\pi,r}\left(m,\pi,r^i\right)$, we make a simplification: we consider the effective action for the homogenous Goldstone mode $\pi_0$ (so with vanishing angular momentum on the sphere $l=0$). This is the mode which is guaranteed to exist by Goldstone's theorem in the $f \rightarrow \infty$ limit. The mode $\pi_0$ depends only on time, so only time derivatives act on it non-trivially. By an effective action for $\pi_0$, we mean that one can consider an effective action for $\pi$, and then derive from it equations of motion and study the solutions for the zero mode $\pi_0$. We can therefore directly work with a Lagrangian which depends only on this mode. 
This Lagrangian takes the form
\be
{\cal L}_{\pi_0,r}\left(m,\pi_0,r^i\right) = \frac12 \alpha\dot{\pi}_0^2 + g_1(r) \left( m\frac{\dot{\pi}_0}{f} \right) + g_2(r) \left( \frac{m\dot{\pi}_0 }{f}\right)^2  + ... + {\cal L}^{(2)}_r \;.
\label{getnga}
\ee
Here $\alpha$ is some arbitrary constant (independent of $f$), the $g_i(r)$ are arbitrary functionals of the fields $r^i$, and ${\cal L}_r^{(2)}$ includes terms which depend on the $r^i$ but do not couple to $\pi_0$. We assume that its leading behaviour is at most second order in time derivatives. The powers of $m$ in the expansion of derivatives follow from the fact that the Lagrangian, in the large $f$ regime, must be a function of the combination 
\be
\left(\partial_{t} \left( m t + \frac{\pi_0}{f} \right)\right)^2 = m^2 + 2 m\frac{\dot{\pi}_0}{f} + \frac{\dot{\pi}^2_0}{f^2}\;,
\label{mfcomb}
\ee
so that each derivative of $\pi_0$ comes with a power of $m$.

We note that in order for terms to remain relevant in the $f \rightarrow \infty$ limit, we need powers of $f$ inside the $g_i(r)$. Since $f$ has dimension $\frac{d-2}{2}$, we see that we can at most include a factor of $f^2$ in the $g_i(r)$. But $f^2$ would saturate the dimensions, and therefore imply there are no fields $r^i$ in the $g_i(r)$, so we can only have one power of $f$. Assuming that the $r^i$ are weakly-coupled, we therefore have the unique possibility
\be
{\cal L}_{\pi_0,r} = \frac12\alpha\dot{\pi}_0^2 + \sum_i \hat{g}_i r^i \left(  m\frac{\dot{\pi}_0}{f} \right) + {\cal L}^{(2)}_r + ...\;,
\ee
where the $\hat{g}_i$ are constants, and we dropped terms sub-leading as $f \rightarrow \infty$. Since $\pi_0$ couples to only one linear combination of the fields $r^i$, we can call this combination $r$. Further, the homogenous Goldstone mode can only source the homogeneous mode if $r$, denoted $r_0$. We can therefore work with the following one-dimensional effective Lagrangian
\be
{\cal L}_{\pi_0,r_0} = \frac12\alpha \dot{\pi}_0^2 + 2 m \beta r_0 \dot{\pi}_0 + \frac12 \dot{r}_0^2 - \frac12\gamma r_0^2 \;,
\label{Lepir0}
\ee
where $\beta$ is some order-one coefficient (so independent of $f$), and $\gamma$ is a constant associated to the mass of $r_0$. 

We note here a subtlety: we have considered the $f \rightarrow \infty$ limit, but the large $f$ regime only requires $f \gg 1$. An issue could arise if there are parametrically large parameters in the theory which appear in the $g_i$ and could spoil the $f$ expansion. We expect that this is not possible, but do not have a general proof. 

In the case where there is a mass hierarchy $\gamma \rightarrow \infty$, we can integrate out $r_0$ to reach an effective theory which has a derivative expansion, as in section \ref{sec:gbg}. However, now we are allowing for small $\gamma$, so that $r_0$ is light. To determine ${\cal L}_{\mathrm{eff}}(m)$ we need to extract the $m$ dependence from $r_0$, evaluated on the solution $\dot{\pi}_0=\dot{r}_0=0$. As before, we do this by relating it to the $\dot{\pi}_0$ dependence in an effective theory ${\cal L}_{\mathrm{eff}}(m,\pi)$ after integrating out $r_0$. Since $r_0$ can be light, this theory does not have a controlled derivative expansion. Nonetheless, it still as an expansion in powers of $\pi$ due to the large $f$ regime. Further, it must still be a function of 
\be
Y = m + \frac{\dot{\pi}_0}{f} \;,
\ee 
and any derivatives acting on it. So we can write, keeping only terms quadratic in $\pi_0$, that
\be
{\cal L}_{\mathrm{eff}}(m,\pi_0) \simeq \frac12 \frac{\partial^2{\cal L}_{\mathrm{eff}}\left(m\right)}{\partial m^2} \left(\frac{\dot{\pi}_0}{f} \right)^2 + \frac{\dot{\pi}_0}{f} \sum_{n=1}^{\infty} a_n(m)\;\partial_t^n \left(\frac{\dot{\pi}_0}{f} \right)  + {\cal L}_{\mathrm{eff}}(m)\;,
\ee
where the $a_n(m)$ are arbitrary coefficients. We can therefore still read off  $\frac{\partial^2{\cal L}_{\mathrm{eff}}\left(m\right)}{\partial m^2}$ from the coefficient of the $\dot{\pi}_0^2$ term. 

In the two-field theory (\ref{Lepir0}), we need to integrate out $r_0$ to reach ${\cal L}_{\mathrm{eff}}(m,\pi_0)$. The equation of motion for the field $r_0$ reads
\bea
-\ddot{r}_0 + 2m\beta \dot{\pi}_0 - \gamma r_0 &=& 0 \label{eqmr0} \;.
\eea
We can solve this as
\be
r_0 = \frac{2 m \beta }{\gamma} \dot{\pi}_0 -2 m \beta  \sum_{n=2}^{\infty} \frac{\partial_t^{2n-1}\pi_0}{\gamma^n} (-1)^n \;.
\label{r0soluallt}
\ee
The key point is that only the first term in (\ref{r0soluallt}) will contribute to the $\dot{\pi}_0^2$ term in ${\cal L}_{\mathrm{eff}}(m,\pi_0)$. So we can substitute 
\be
r_0 \rightarrow \frac{2 m \beta}{\gamma} \dot{\pi}_0 \;,
\ee
into (\ref{Lepir0}), which yields
\be
\frac{1}{f^2}\frac{\partial^2{\cal L}_{\mathrm{eff}}\left(m\right)}{\partial m^2} = \alpha + \frac{\left(2\beta m\right)^2}{ \gamma} \;.
\label{d2lmfgen}
\ee

We now constrain the parameters $\alpha$, $\beta$, $\gamma$, $m$ and $f$ within the two-field theory. The full equations of motion for the action (\ref{Lepir0}) read
\bea
\alpha \ddot{\pi}_0 + 2 m \beta \dot{r}_0 &=& 0 \;, \label{eqm1}\\
-\ddot{r}_0 + 2m\beta \dot{\pi}_0 - \gamma r_0 &=& 0 \label{eqm2} \;.
\eea
We can solve (\ref{eqm1}), and insert it into (\ref{eqm2}), to yield 
\be
\ddot{r}_0 + r_0 \left( \frac{\left(2m\beta\right)^2}{\alpha}  + \gamma  \right) =0 \;.
\ee
This leads to an unstable mode unless 
\be
\frac{\left(2m\beta\right)^2}{\alpha}  + \gamma > 0 \;.
\label{massconst}
\ee
Now we note that $\alpha > 0$ by positivity of the kinetic terms.\footnote{Note that, by adding momentum modes to $\pi$, we can also extract the relation $\alpha =  \frac{1}{m}\frac{\partial{\cal L}_{\mathrm{eff}}\left(m\right)}{\partial m} = \frac{Q}{\mathrm{Vol}_{d-1}m}>0$.} Let us also assume that the mass parameter is positive $\gamma > 0$. Then, we can write (\ref{massconst}) precisely as the combination in (\ref{d2lmfgen}), yielding 
\be
\frac{\partial^2{\cal L}_{\mathrm{eff}}\left(m\right)}{\partial m^2} > 0 \;.
\ee
We therefore find again that convexity is implied by stability. Note, however, in this case we find a weaker result than in section (\ref{sec:gbg}) because we only have a stability problem rather than a negative kinetic term. We recover the results of that section in the $\gamma \rightarrow \infty$ limit, where then convexity is directly related to the sign of $\alpha$.

Let us now return to the assumption $\gamma >0$. First, we note that if $\gamma$ is large in magnitude and negative, then we obtain an instability. So our matching of instability to convexity is problematic only if $\gamma$ is in the range $0 >\gamma > -\frac{\left(2m\beta\right)^2}{\alpha}$. In fact, $\gamma$ is bounded from below by a parameter which is of order one, basically the sphere momentum scale. The reason is that if there are momentum modes whose momentum contribution to the dispersion relation is less than that of the negative $\gamma$, there will be an instability (the usual one for infrared momentum modes in flat space). It is the quantisation of momentum modes which allows for a negative $\gamma$. In flat space, $\gamma$ would have to be positive for stability of the theory. On the sphere $\gamma$ receives a further positive contribution from the sphere curvature, relative to the flat space value, so should become more positive. However, this does not prove $\gamma > 0$, since it need not be that the state is stable in flat space, only on the sphere. We therefore do not have a proof of $\gamma > 0$. It is possible to show, however, that if it is negative, then the concavity would be bounded from below $\frac{\partial^2{\cal L}_{\mathrm{eff}}\left(m\right)}{\partial m^2} < -\alpha$. So there is no continuous interpolation between convex spectra and concave ones as we vary $\gamma$ across $\gamma=0$. This (infinite jump) discontinuity at $\gamma=0$ further suggests that it must be positive. 

In this analysis we restricted only to the homogenous modes. It is possible to perform an analysis including also the momentum modes. We do this, though in a slightly more restricted theory, in appendix \ref{sec:two-field-stability}. In particular, this way one recovers again also the constraint (\ref{sublum}).

%%%%%%%%%%%%%%%%%%%%%%%%%%%%%%%%%%%%%%%%%%%%%%%%%
\section{Convexity in charge}
\label{sec:cco}
%%%%%%%%%%%%%%%%%%%%%%%%%%%%%%%%%%%%%%%%%%%%%%%%%

The results of section \ref{sec:effgb} can be summarised concisely: in the large $f$ regime (\ref{largef}), we have an effective theory for a Goldstone boson $\pi$ about a charged state, with chemical potential $m$. Consistency of this theory, specifically positive kinetic terms and stability, requires that the Lagrangian must be a convex function of $m$, so
\be
\frac{\partial^2{\cal L}_{\mathrm{eff}}\left(m\right)}{\partial m^2}  > 0 \;.
\label{convLm}
\ee
Here, ${\cal L}_{\mathrm{eff}}\left(m\right)$ denotes the Lagrangian in which all the $m$ dependence has been made explicit, by integrating out the relevant fields. 

We are interested in the spectrum of the dimensions of charged operators $\Delta(Q)$. By the state-operator correspondence, this is given by the energies of the dual states
\be
\Delta\left(Q\right) = E\left(Q\right) R \;.
\ee
Here, by $E(Q)$, we denote the expectation value of the Hamiltonian operator $\hat{H}$ on a charge eigenstate of charge $Q$, so
\be
E(Q) = \langle Q|\hat{H}|Q\rangle \;.
\ee
In particular, we are interested in the dimensions of the lowest dimension operators of charge $Q$, and so the energy is that of the lowest energy state of charge $Q$.

Classically, the Lagrangian as a function of $m$ is the Legendre dual of the Hamiltonian as a function of charge $Q$. So we have
\be
\mathrm{Classical\;:\;} H\left(Q\right) = m \;Q - L(m) \;,\;\; Q = \frac{\partial L\left(m\right)}{\partial m} \;.
   \label{Legendtrans}
\ee
A Legendre transform preserves convexity, indeed the Legendre dual is sometimes referred to as the convex conjugate.\footnote{It is worth noting a subtlety at this point, which is that the Legendre transform (\ref{Legendtrans}) is not well-defined if there are multiple values of $m$ which give the same $Q$. However, this does not happen if the convexity condition (\ref{convLm}) is satisfied. This is because we are restricting to positive charges $Q > 0$, and so both the first and second derivatives of ${\cal L}\left(m\right)$ are positive and therefore it is a monotonic function.} Therefore, convexity of $L(m)$ in $m$, as in (\ref{convLm}), implies convexity of $H(Q)$ in $Q$, and therefore also of $\Delta(Q)$ in $Q$. This is the required result.

There are a number of subtleties with this argument. The first is that since our analysis in section \ref{sec:ccp} restricted to homogeneous states, we are therefore restricted to scalar operators. This does not mean that there are only scalars in the \textsc{cft}, but that the lowest dimension operators of a given charge are scalar in nature (they may be formed from scalar combinations of fermions for example). So we assume that the \textsc{cft} operator spectrum respects this scalar property, at least in the charge range of interest. We leave the extension of our analysis to non-scalar operators for future work.

The second important point is that in order to utilise the classical relation between the Hamiltonian and Lagrangian, we need a semi-classical saddle point of the path integral. In the large $f$ regime (\ref{largef}), there is a weakly-coupled (quadratic) effective theory for the Goldstone boson. This therefore implies a valid semi-classical treatment. 

The third point is possibly the most important. The result (\ref{convLm}) holds when the second derivative is evaluated about a given value of $m$. More precisely, it arises from an effective action for a Goldstone boson about a given charged state. We are after understanding convexity of $\Delta(Q)$, so convexity of {\it different} states. Each state of a given charge $Q$ will have its own Goldstone boson, and it can be that different charges have different Goldstone bosons. In that case, we cannot use the effective action for one of the Goldstone bosons to study the other charged states. What our results show is that if we have some set of operators (of lowest dimension for their charge) within some charge or parameter range, such that the dual states to them all have the same Goldstone boson (assuming they lie in the large $f$ regime), then that set of operators is convex. We describe this in more detail in section \ref{sec:hamtopath} below.

We see also a possible way to avoid convexity in theories: if we have a situation where the lowest energy states for the different charges have different Goldstone bosons, then there is no way to relate them through our analysis. This is precisely what is realised in the counter-example to convexity discovered in \cite{Sharon:2023drx}. We present an analysis of a toy-example of such a case in section \ref{sec:simsce}. Nonetheless, in these theories we still have families of (lowest dimension) operators convex in charge, but they may have parametrically large charge. 

When a set of states, or operators, shares a Goldstone boson or not is not simple to deduce in general. For weakly-coupled theories it is simple: if we have a charged operator $\Phi$, then $\Phi^n$ all share the same Goldstone boson. 

%%%%%%%%%%%%%%%%%%%%%%%%%%%%%%%%%%%%%%%%%%%%%%%%%%%%%%%%%%
\subsection{Hamiltonian to Lagrangian mapping}
\label{sec:hamtopath}
%%%%%%%%%%%%%%%%%%%%%%%%%%%%%%%%%%%%%%%%%%%%%%%%%%%%%%%%%%

In this section we discuss in more detail the mapping between the Hamiltonian as a function of charge to the Lagrangian as a function of the chemical potential. 

The Goldstone boson $\pi$ is defined over a fixed charge state. To relate different charged states, we must have some theory that captures those states. We must then embed the Goldstone boson into the degrees of freedom of that theory. So let us define a charge range ${\cal M}$, over which the theory is well described by some relevant degrees of freedom, parameterised with an index $i$, and denoted as $\phi^{\cal M}_i$. So, for example, ${\cal M}$ could be the large charge regime, or it could be a weakly-coupled regime, or any other charge range where the theory is well described by the same set of degrees of freedom. 

Within the domain ${\cal M}$, we may have a number of different charged states that are in the large $f$ regime, and so are described by some semi-classical effective theory involving a Goldstone boson $\pi$. But the effective theory, and even the Goldstone boson itself, about each such state can be different.  

Let us introduce some notation to classify these possible effective theories and Goldstone bosons. We would like to fix the embedding of the Goldstone boson into the degrees of freedom $\phi^{\cal M}_i$. Let us denote the combinations of the $\phi^{\cal M}_i$ that realise the Goldstone boson around a given charged state as $\chi$. More precisely, the Goldstone boson appears embedded into the $\phi^{\cal M}_i$ only through the combination
\be
\chi = m t + \frac{\pi}{f} \;.
\label{chilam}
\ee 
Therefore, what $\chi$ captures is the relation between the Goldstone boson $\pi$ and the symmetry breaking expectation value $m t$. In a theory, we have one $U(1)$ and so one combination of the $\phi^{\cal M}_i$ which transforms under it. However, in general, not all the fields charged under the $U(1)$ need to develop an expectation value $mt$. The combination which develops this expectation value, on a given state, is what we denote by $\chi$, and what we call the embedding of the Goldstone boson. 

As a quick informative example one can consider a theory with two complex scalars $\phi_1$ and $\phi_2$, both of which transform under the $U(1)$ global symmetry. Then if only $\phi_1$ develops an expectation value in the minimal energy charged state, then $\chi$ and the Goldstone boson are given by the phase of $\phi_1$. Similarly, if only $\phi_2$ develops the expectation values, then the Goldstone boson is embedded into its phase. 

We will always consider the lowest energy state for a given charge $Q$. If the embedding remains constant over those states, then varying $m$ in (\ref{chilam}) corresponds to different charge states, and we say that all these states {\it share the same Goldstone boson}. We now see in what sense we can have convexity with respect to different states: convexity in $m$ for an effective theory for a Goldstone boson implies convexity in states which share the same Goldstone boson. 

Let us make this more precise. To map the Lagrangian analysis of the Goldstone effective theory to the energy of the states we need to consider the path integral representation for the Hamiltonian expectation value. Working with path integrals, it is useful to rotate to Euclidean time. We denote Minkowski time by $t$ and Euclidean time by $\tau$, with 
   \be
   \tau = it \;.
   \ee
The Minkowski and Euclidean actions are then
\be
S_E = \int {\cal L}_E\left(\tau\right) d^{d-1}x d\tau \;,\;\; S = \int {\cal L}(t)\; d^{d-1}x dt \;,
\ee
and the associated Lagrangians are
\be
{\cal L}_E\left(\tau\right) = - {\cal L}\left(t \rightarrow -i\tau\right) \;.
\label{eutomink}
\ee

It will also be informative to work at finite temperature. So we take the spacetime of the form $S^1 \times S^{d-1}$, with the circumference of the $S^1$ denotes as $\beta$. So we have that $\int d\tau = \beta$. The zero temperature limit is $\beta \rightarrow \infty$. 

Let us consider the energy of some state of charge $Q$.\footnote{Note that this part is a generalisation of the analysis in \cite{Badel:2019oxl}.} The state is arbitrary for now, and is not necessarily an energy eigenstate or dual to any fixed dimension operator. The lowest energy state of charge $Q$ will, in the large $f$ regime, have some embedding of the Goldstone boson denoted by $\chi$. We let the general state of charge $Q$ have the same Goldstone embedding, so we denote it as $|\psi_Q,\chi\rangle$. In mapping to a path integral, we need to present some representation (wavefunction) for the state in a basis of the fields appropriate to the path integral, so the $\phi^{\cal M}_i$. The choice of embedding is then concretely taking a state such that 
\be
\langle \phi^{\cal M}_i |\psi_Q,\chi,\tau\rangle_S \sim e^{i Q \chi_0} \;,
\label{defchemgl}
\ee 
where $\chi_0$ is the zero (homogeneous) mode of $\chi$, and $\langle \phi^{\cal M}_i|$ are the basis of states where the fields take the definite values $\phi^{\cal M}_i$. Here we made the Euclidean time $\tau$ dependence manifest, and added a subscript $S$, to denote that we are in the Schrodinger picture. Note that we can write the zero mode as
\be
\chi_0 = \frac{1}{\mathrm{V}_{d-1}}\int_{S^{d-1}} \chi \;d^{d-1}x\;.
\ee
The state we are after, preserving the superfluid symmetries, is one where all the fields are constant in space and in time, apart from $\chi_0$ (which has a linear time dependence interconnected with the charge). Let us write this state explicitly, though now in the Heisenberg picture \cite{Badel:2019oxl}
\be
|\psi_Q,\chi\rangle_H = \int {\cal D} \alpha e^{i \frac{Q}{\mathrm{V}_{d-1}} \int_{S^{d-1}} \alpha \;d^{d-1}x } | \alpha\rangle \;,
\label{wavefunchl}
\ee
where $| \alpha\rangle$ is a state in which $\chi$ can have an arbitrary (fixed) spatial profile $\chi({\bf x})=\alpha({\bf x})$, but all the other fields are (fixed) spatially constant. 

We now consider the time evolution from $\tau_i=-\frac{\beta}{2}$ to $\tau_f=+\frac{\beta}{2}$:
\be
\langle \psi_Q,\chi | e^{-\beta H } | \psi_Q,\chi  \rangle = {\cal Z}^{-1} \int {\cal D} \chi_i {\cal D} \chi_f e^{-i \frac{Q}{\mathrm{V}_{d-1}} \int \left( \chi_f - \chi_i \right) d^{d-1}x} \int_{\chi_i}^{\chi_f} {\cal D} \chi e^{-S_E} \;.
\label{qhqpi}
\ee
Note that by ${\cal D}\chi$ we denote the path integration over $\chi$, but also over any other fields in the theory. The boundary conditions of the path integral are such that all other fields apart from $\chi$ are constant in spacetime. 
The normalisation factor is
\be
{\cal Z} = \int {\cal D} \chi e^{-S_E} \;.
\ee
By writing
\be
\int \left( \chi_f - \chi_i \right) d^{d-1}x = \int_{-\frac{\beta}{2}}^{\frac{\beta}{2}} d\tau \int \dot{\chi} \; d^{d-1}x \;,
\ee
we can write 
\be
\langle \psi_Q,\chi | e^{-\beta H } | \psi_Q,\chi \rangle = {\cal Z}^{-1}\int {\cal D} \chi e^{-S_{E,\mathrm{eff}}} \;,
\label{wavefunchlshort}
\ee
where 
\be
S_{E,\mathrm{eff}} = S_E + \int i \frac{Q}{\mathrm{V}_{d-1}} \dot{\chi} \;d\tau d^{d-1}x  \;.
\label{effacbtr}
\ee

Equations (\ref{wavefunchlshort}) and (\ref{effacbtr}) contain the physics of interest. We can decompose $| \psi_Q,\chi  \rangle $ into energy eigenstates as
\be
| \psi_Q,\chi \rangle = \sum_i \;\langle Q, E_i | \psi_Q,\chi  \rangle \;|  Q, E_i \rangle \;,
\ee
where $|  Q, E_i \rangle$ is an energy eigenstate of energy $E_i$ and charge $Q$. The left-hand-side of (\ref{wavefunchlshort}) then reads
\be
\langle \psi_Q,\chi | e^{-\beta H } | \psi_Q,\chi  \rangle = \sum_i \left|\langle Q, E_i | \psi_Q,\chi \rangle\right|^2 e^{-\beta E_i} \;.
\label{sumeees}
\ee
The right-hand-side of (\ref{wavefunchlshort}) will be dominated by saddle points of the path integral, at least while we are in the large $f$ regime. 
In the zero temperature limit $\beta \rightarrow \infty$, the sum (\ref{sumeees}) will be dominated by the lowest energy state, so we can write
\be
\lim_{\beta \rightarrow \infty}  \langle \psi_Q,\chi | e^{-\beta H } | \psi_Q,\chi \rangle= \left|\langle Q, E_0 | \psi_Q,\chi  \rangle\right|^2 e^{-\beta E_0} \;,
\ee
where $E_0$ is the energy of the lowest energy eigenstate of charge $Q$. Similarly, the path integral will be dominated by the smallest action saddle.\footnote{Note that this saddle is at imaginary values of $\chi$ when we take it as a function of Euclidean time $\tau$: $\chi = -i m \tau$. This requires the usual analytical continuation of the fields in the path integral.} We can perform a saddle point approximation to calculate the classical equations of motion. We write the effective Lagrangian as
\be
L_{E,eff}  = L_E + i Q \dot{\chi} \;.
\label{leff}
\ee 
The equations of motion have a bulk and boundary term contribution. The boundary variation gives
\be
\frac{\partial L_E}{\partial \dot{\chi}_0} \left( \tau_i \right) = \frac{\partial L_E}{\partial \dot{\chi}_0} \left( \tau_f \right) = - i Q \;.
\label{bountxl}
\ee
The bulk equations are just the current conservation equation for $\chi$. 

The classical solutions we are interested in have $\dot{\chi}$, and all the other fields, constant in time and space. In order to have this, we need to match the boundary conditions (\ref{bountxl}) to the bulk equations. This is why we took the Goldstone embedding of the general state of charge $Q$, to match that of the lowest energy one. We also need to choose the values for all the other fields in $| \psi_Q,\chi  \rangle$ above equal to their values on the saddle point.
In that case, we can solve the saddle point equations, or the equations of motion, with profiles constant in time, and we have the boundary terms giving
\be
\frac{\partial L_E}{\partial \dot{\chi}_0} = i\frac{\partial L_E}{\partial m} = -i\frac{\partial L}{\partial m} = -i Q \;,
\ee
with $L$ the Minkowski signature Lagrangian. We therefore recover the Minkowski constraint for the Legendre transform
\be
\frac{\partial L}{\partial m} = Q \;.
\label{piltq}
\ee
Further, since all the fields are constant in time in the profile about the saddle, we can perform the time and space integration so that we obtain
\be
\beta E_0 = \beta \left( L_E + Q m \right) = \beta \left( -L + Q m \right) \;.
\ee
This energy is the same as calculated through the Hamiltonian
\be
H = -L + Q \;m  \;.
\label{pilth}
\ee
Together, (\ref{piltq}) and (\ref{pilth}) give the classical Legendre transform. This is the result we are after since a Legendre transform preserves convexity.

%%%%%%%%%%%%%%%%%%%%%%%%%%%%%%%%%%%%%%%%%%%%%%%%%%%%%%%%%%
\subsection{Simple scenario with different Goldstone embeddings}
\label{sec:simsce}
%%%%%%%%%%%%%%%%%%%%%%%%%%%%%%%%%%%%%%%%%%%%%%%%%%%%%%%%%%

It is informative to illustrate the discussion on the Goldstone embedding with a simple scenario. Consider a \textsc{cft} which is weakly-coupled, and has two complex fields $\phi_1$ and $\phi_2$, such that their charges under the $U(1)$ global symmetry are\footnote{A simple way to construct such a theory could be to take two copies of the $O(2)$ model, as studied in section \ref{sec:to2m}, in $d=4-\epsilon$ dimensions and add an interaction term $\lambda \phi_1^3 \phi_2$. We thank Adar Sharon and Masataka Watanabe for brining this possibility to our attention.}
\be
Q_{\phi_1} = 1 \;,\;\; Q_{\phi_2} = -3 \;.
\ee
Such a theory would not have a convex spectrum of operators (of lowest dimension for a given charge). For example, consider the lowest dimension operators of charges around $Q=3N$, with $N$ an integer that can be large, but still in the weakly-coupled regime:
\bea
Q=3N \;&:&\; \left(\phi_2^*\right)^N \;, \\
Q=3N+1 \;&:&\; \left(\phi_2^*\right)^N \phi_1 \;, \\
Q=3N+2 \;&:&\; \left(\phi_2^*\right)^N \phi_1^2 \;, \\
Q=3N+3 \;&:&\; \left(\phi_2^*\right)^{N+1} \;.
\eea
Since the theory is weakly coupled, it is clear that this spectrum is not convex
\bea
\Delta\left(3N+2\right) &=& \Delta\left(\left(\phi_2^*\right)^N \phi_1^2\right) \sim N+2 \\
\Delta\left(3N+3\right) &=& \Delta\left(\left(\phi_2^*\right)^{N+1}\right) \sim N+1 \;,
\eea
and so $\Delta\left(3N+3\right) < \Delta\left(3N+2\right)$, which violates convexity.

In this scenario, the absence of convexity can be understood in terms of the variation of the Goldstone embedding, for the lowest energy state of a given charge.
Let us write the fields as
\be
\phi_1 = \frac{1}{\sqrt{2}} a^{(1)} e^{i \chi^{(1)}} \;,\;\; \phi_2 = \frac{1}{\sqrt{2}} a^{(2)} e^{i \chi^{(2)}} \;.
\ee
The choice of states $| \psi_{Q}, \chi^{\Lambda}, \tau \rangle_S$, which defines the Goldstone embedding, must be matched onto the states dual to the associated operators. The wavefunctions of those (lowest energy) states transform as 
\bea
Q=3n \;&:&\; \langle a^{(1)},\chi^{(1)},a^{(2)},\chi^{(2)}|\left(\phi_2^*\right)^n(\tau) | 0 \rangle \sim e^{i Q \left(-\frac13 \chi^{(2)}_0 \right) }\;, \nn \\
Q=3n+1 \;&:&\; \langle a^{(1)},\chi^{(1)},a^{(2)},\chi^{(2)}|\left(\phi_2^*\right)^n(\tau) \phi_1(\tau) | 0 \rangle \sim e^{i \left(-n \chi^{(2)}_0 + \chi^{(1)}_0\right) } \;, \nn \\
Q=3n+2 \;&:&\; \langle a^{(1)},\chi^{(1)},a^{(2)},\chi^{(2)}|\left(\phi_2^*\right)^n(\tau) \phi_1^2(\tau) | 0 \rangle \sim e^{i  \left(-n \chi^{(2)}_0 + 2 \chi^{(1)}_0 \right) }\;, \nn \\
Q=3n+3 \;&:&\; \langle a^{(1)},\chi^{(1)},a^{(2)},\chi^{(2)}|\left(\phi_2^*\right)^{n+1}(\tau) | 0 \rangle \sim e^{i Q \left(-\frac13 \chi^{(2)}_0 \right) }\;.
\eea
We therefore see that the only family of lowest energy states which have the same Goldstone embedding is $\left(\phi_2^*\right)^n$, which have charges $Q=3n$ with $n \in \mathbb{N}$. For this family we can take 
\be
| \psi_{Q}, \chi^0, \tau \rangle_S \sim \left(\phi_2^*\right)^{\frac13 Q}(\tau) | 0 \rangle \;,
\ee
so that
\be
\langle a^{(1)},\chi^{(1)},a^{(2)},\chi^{(2)} | \psi_{Q}, \chi^0, \tau \rangle_S \sim e^{i Q \left(-\frac13 \chi^{(2)}_0 \right) }\;,
\label{2ndemb}
\ee
and therefore $\chi = -\frac13 \chi^{(2)}$.
Our results then imply that this family, $\left(\phi_2^*\right)^n$, is convex. 

In this model, convexity in terms of (\ref{ccc}) holds for $q_0=3$. By taking a parametrically large number of copies of this model, it is possible to make $q_0$ parametrically large \cite{Sharon:2023drx}.

%%%%%%%%%%%%%%%%%%%%%%%%%%%%%%%%%%%%%%%%%%%%%%%%%
\section{Example: the $O(2)$ model}
\label{sec:to2m}
%%%%%%%%%%%%%%%%%%%%%%%%%%%%%%%%%%%%%%%%%%%%%%%%%

In this section we study an example model. The example we consider is actually a general class of theories, where we assume that after integrating out heavy modes, one is left with a single complex scalar which realises the $U(1)$ symmetry as a phase rotation. It is a generalisation of the $O(2)$ model studied in \cite{Badel:2019oxl}, and technically the calculation is the same as the one in \cite{Badel:2019oxl}. 

Consider a weakly-coupled model for a \textsc{cft}, written in terms of a complex field \(\phi = \frac{1}{\sqrt{2}}a e^{i \chi}\) that transforms linearly under a \(U(1)\) symmetry. 
	If \(a \neq 0\), we can write the (Minkowski signature) Lagrangian in an (approximately) scale-invariant form as an expansion in the coupling
   \begin{equation}
     {\cal L} = \frac{1}{2} \partial_{\mu} a \partial^{\mu} a + \frac{1}{2} a^2 \partial_{\mu} \chi \partial^{\mu} \chi - \frac{(d-2) {\cal R} }{8(d-1)} a^2 - \frac{d-2}{8d} \;g\; a^{\frac{2d}{d-2}} + \text{sub-leading in \(g\),}
     \label{simmdac}
   \end{equation}
   where $g$ is a perturbative coupling. We consider the theory on $\mathbb{R} \times S^{d-1}$, where the sphere has an associated radius $R$, and a curvature ${\cal R}$ which contributes to the Lagrangian through a quadratic term in $a$.  
   
   The simplest examples are the Wilson--Fisher point for the \(O(2)\) model in \(d = 4-\epsilon\) dimensions where
   \begin{equation}
     g_{4-\epsilon} = \frac{\left(4 \pi\right)^{2} \epsilon}{5} + {\cal O}\left(\epsilon^2\right) \; ,
   \end{equation}
   the critical point of the \(O(N)\) vector model in \(d=3\) dimensions at large \(N\) where~\cite{Appelquist:1982vd,Alvarez-Gaume:2019biu,Orlando:2021usz}
   \begin{equation}
     g_{N} = \frac{8\left(4\pi\right)^2}{N^2} + {\cal O}\left(1/N^3\right) \; ,
   \end{equation}
   or the critical point of the \(SO(N)\) NJL model in \(d=3\) dimensions at large \(N\), where~\cite{Dondi:2022zna}
   \begin{equation}
     g_{N} = \frac{8\left(4\pi\right)^2}{\kappa_0^6 N^2} + {\cal O}\left(1/N^3\right) \; , \hspace{2em} \kappa_0^6 = 2.98119\dots
   \end{equation}
   Note that in the last two cases, the theory is strongly coupled near the fixed point (and the NJL model is actually a fermionic model), but the physics around the fixed-charge state is written in terms the of the bosonic field $\chi$ which is effectively weakly-coupled at large $N$. Similarly, we expect matrix-type theories to be controlled by the appropriate 't Hooft coupling.\footnote{A simple example is the asymptotically safe theory of~\cite{Litim:2014uca}, which has a matrix scalar sector which is controlled by $\epsilon/N_f^2$ with $\epsilon = N_f/N_c - 11/2$ ~\cite{Orlando:2019hte}.} 
    
The theory has two large $f$ regimes, described in table \ref{tab:fmQso2}. In the regime $g Q \gg 1$, the large charge regime, the radial mode $a$ becomes very massive and can be integrated out to yield an effective theory as in section \ref{sec:gbg}. The other large $f$ regime is $g Q \ll 1$ and $Q \gg 1$. In that case, there is no gap to the radial mode mass and we are in the scenario of section \ref{sec:gbl}. We consider these two settings in turn. 

First, there is some analysis which can done irrespective of the charge regime.
For simplicity, we restrict to the $d=4$ case, which sets ${\cal R}=6/R^2$ and $V = 2 \pi^2 R^3$. In this case the action is
\be
 {\cal L} = \frac{1}{2} \partial_{\mu} a \partial^{\mu} a + \frac{1}{2} a^2 \partial_{\mu} \chi \partial^{\mu} \chi - \frac12 \frac{a^2}{R^2} - \frac{1}{16} \;g\; a^{4} \;.
     \label{simmdac4d}
\ee
We are looking for a saddle, so a solution to the equations of motion, where $a$ is a constant in spacetime, and $\chi$ takes the form
\be
\chi = m t \;.
\label{backmt}
\ee
The classical equations of motion for the field $a$ give
  \be
  \langle a \rangle^2 = \frac{4\left(m^2 - 1/R^2\right)}{g} \;.
  \label{perta}
  \ee
  Note that this only makes sense over the domain
  \be
  m^2 R^2 \geq 1 \;.
  \label{adom}
  \ee
 Substituting the solution for $a$, so evaluating the action on the saddle, yields the effective Lagrangian
 \be
 {\cal L}_{\mathrm{eff}}(m) = \frac{\left(m^2 - 1/R^2\right)^2}{g} \;.
 \ee
 We note that it is indeed convex. More precisely, 
 \be
 \frac{\partial^2}{\partial m^2} {\cal L}_{\mathrm{eff}}(m) = \frac{4}{g} \left(3m^2 - 1/R^2\right)\;,
 \label{pard2ml}
 \ee
 and therefore it is convex over the domain, 
 \be
 \text{Convex domain:}\;m^2 R^2 > \frac{1}{3} \;,
 \label{convexdom}
 \ee
 which is automatically implied by~(\ref{adom}).
 
 We are interested in the domain $Q \geq 0$, with the charge $Q$ defined as
 \be
 Q \equiv \left(2 \pi^2 R^3 \right) \frac{\partial {\cal L}_{\mathrm{eff}}(m)}{\partial m} = \frac{ 8\pi^2 R^3  m \left(m^2-1/R^2\right)}{g} \;.
 \label{Qasm}
 \ee
 %We have, for convenience, flipped the sign of the charge in the definition so that it is positive for $m^2 > 1$. 
 We see that the domain $Q\geq 0$ is also implied by~(\ref{adom}), and the restriction $m\geq 0$. 
Inverting~(\ref{Qasm}) we have
 \be
 m R = \frac{3^{\frac13}+\left(9 \frac{g Q}{\left(4\pi\right)^2} - \sqrt{81 \left(\frac{g Q}{\left(4\pi\right)^2}\right)^2 -3}\right)^{\frac23}}{3^{\frac23}\left(9 \frac{g Q}{\left(4\pi\right)^2} - \sqrt{81 \left(\frac{g Q}{\left(4\pi\right)^2}\right)^2 -3}\right)^{\frac13}} \;.
 \label{fullm}
 \ee
We can now study the two large $f$ regimes separately.

\subsection{Large $f$ regime with $gQ \gg 1$}

 In the $gQ \gg 1$ regime, we have that (\ref{fullm}) is approximated as
 \be
 m R \simeq \left(\frac{gQ}{8\pi^2}\right)^{\frac13} \;.
 \ee
In terms of the Goldstone boson, we can read off $f$ as
\be
f^2 = \langle a \rangle^2 \simeq \frac{4}{gR^2} \left(\frac{gQ}{8\pi^2}\right)^{\frac23} \;.
\ee
So we see that we are in the large $f$ regime. 

The other parameter we are interested in is the radial mode mass. Indeed, it is informative to obtain the effective theory for the homogenous Goldstone boson $\pi_0$ and radial $r_0$ modes. Note that the homogenous Goldstone zero mode only sources the homogenous radial mode, so it is consistent to restrict to them. We set $R=1$ henceforth, and reinstate only when informative. We expand out
\be
a = f + r_0 \;,\;\; \chi = m t + \frac{\pi_0}{f} \;.
\ee
The leading terms in the effective action read
\bea
{\cal L} &=& \frac{1}{2}  \dot{r}_0^2 
+ \frac{1}{2} \dot{\pi}_0^2 + 2 m  r_0  \dot{\pi}_0 - \left(m^2-1 \right) r_0^2  \;.
\label{leadingeffac}
\eea
This matches onto the action (\ref{Lepir0}), with 
\be
\alpha=1 \;,\;\;\beta = 1 \;,\;\; \gamma = 2\left(m^2-1 \right) \;.
\label{abgeff}
\ee
Note that we always have $\gamma > 0$.
We can read off the radial mode mass $M_{r_0}$ from the quadratic term for $r_0$, which gives
\be
M_{r_0} \simeq \frac{\sqrt{2}}{R} \left(\frac{gQ}{8\pi^2}\right)^{\frac13} \;.
\ee
This is much heavier than the momentum modes energy $\frac{1}{R}$, and so can be safely integrated out. 
The relevant terms in the equation of motion for $r_0$ read
\be
- \ddot{r}_0 + 2 m \dot{\pi}_0 - \frac12 g f^2 r_0 =0 \;.
\ee
Since the radial mode is very massive, we can integrate it out by restricting $\ddot{r}_0=0$, and setting
\be
r_0 = \frac{4m \dot{\pi}_0}{gf^2} \;.
\label{so2r0eq}
\ee
Plugging this back into the action (\ref{leadingeffac}) the yields the coefficient of the quadratic kinetic term
\be
{\cal L} \supset \frac12 \left(\frac{4\left(3m^2-1\right)}{g} \right)\left(\frac{\dot{\pi}_0}{f}\right)^2 = \frac12 \frac{\partial^2{\cal L}(m)}{\partial m^2}\left(\frac{\dot{\pi}_0}{f}\right)^2 \;,
\ee 
where we used (\ref{pard2ml}). This reproduces the general result (\ref{ltodi}). 

\subsection{Large $f$ regime with $gQ \ll 1$ and $Q \gg 1$}

The other large $f$ regime is for $gQ \ll 1$ and $Q \gg 1$. In this case, we have from (\ref{fullm}) that 
\be
m R \simeq 1 + \frac{g Q}{\left(4\pi\right)^2}\;.
\ee
Therefore, we have from (\ref{perta}) that
\be
f^2 \simeq \frac{Q}{R^2 2\pi^2} \;.
\ee
We see that we are indeed in a large $f$ regime. 

In this case, we do not have a mass gap to the radial mode. We therefore should follow the analysis in section \ref{sec:gbl}. The match of the actions is given by the values (\ref{abgeff}). Putting these into (\ref{d2lmfgen}) then yields (\ref{pard2ml}).

 Note that, because we are at weak coupling, it is possible to match the results onto a calculation done using perturbative Feynman diagrams. 
 We can expand at small coupling
 \be
 m R = 1+ \frac{g Q}{\left(4\pi\right)^2} - \frac32 \left(\frac{g Q}{\left(4\pi\right)^2}\right)^2 + ... \;.
 \label{pertm}
 \ee
 Performing the Legendre transform to the Hamiltonian
 \be
 H_{\mathrm{eff}}\left(Q\right) = m Q - 2 \pi^2 R^3 {\cal L}_{\mathrm{eff}}(m) \;,
 \label{legtramex}
 \ee
 gives
 \be
 H_{\mathrm{eff}}\left(Q\right) = \frac{Q}{R} \left( 1 + \frac12 \frac{gQ}{(4\pi)^2} - \frac12 \left( \frac{gQ}{(4\pi)^2} \right)^2 + \dots \right)\;. 
 \ee
 The state-operator correspondence matches this energy to the operator dimension, through
 \be
  H_{\mathrm{eff}}\left(Q\right) R = \Delta(Q) \;. 
 \ee
 The result matches the perturbative calculation~\cite{Badel:2019oxl,Arias-Tamargo:2019xld} of the operator dimensions
 \be
 \Delta(Q) = Q \left(1 + \frac12 \frac{gQ}{(4\pi)^2} - \frac12\left(\frac{gQ}{(4\pi)^2}\right)^2  + \text{subleading in \(Q\) at each order in \(g\)} \right) \;,
 \ee
 up to corrections sub-leading in $Q$ (at each order in $g$). Note that we have reproduced the one-loop correction to the operator dimension through a leading classical evaluation of the action on the charged state. The sub-leading corrections in $Q$ come from evaluating perturbations about this background \cite{Badel:2019oxl,Antipin:2020abu}.

%%%%%%%%%%%%%%%%%%%%%%%%%%%%%%%%%%%%%%%%%%%%%%%%%
\section{Convexity, expectation values and eigenstates}
\label{sec:cce}
%%%%%%%%%%%%%%%%%%%%%%%%%%%%%%%%%%%%%%%%%%%%%%%%%
	
In this section we discuss a different perspective on convexity. It is a reformulation of convexity as some property of the Hilbert space of the theory. Specifically, we show that convexity of the operator spectrum is implied by the requirement that the lowest energy state for a given {\it expectation value} of the charge operator is a charge eigenstate. 

 There is some sense in which this is similar in spirit to the analysis of section \ref{sec:ccp}. Our results can be interpreted as the statement that we should be able to write an effective theory about the lowest energy state for a given charge expectation value (equal to a charge eigenvalue). If that state is not a charge eigenstate, then such an effective theory would not satisfy basic properties such as cluster decomposition. However, we do not know how to prove that such a good effective theory should exist for the lowest energy state of a given charge expectation value. We do expect such a theory to exist for the lowest energy charge eigenstate, and that is what was demanded in section \ref{sec:ccp}.
 
 There may be other reasons to demand that the lowest energy state of a fixed charge expectation value should be a charge eigenstate. Perhaps it can be shown holographically. We do not know how to prove this requirement, and so in summary this section is a reformulation of convexity which may be useful. 

	%%%%%%%%%%%%%%%%%%%%%%%%%%%%%%%%%%%%%%%%%%%%%%%
	\subsection{Convexity and charge eigenstates in a general QFT}
	\label{sec:conpure}
	%%%%%%%%%%%%%%%%%%%%%%%%%%%%%%%%%%%%%%%%%%%%%%%

	In this section we show that, in a general Quantum Field Theory, convexity in energy of the spectrum of charge eigenstates is implied by requiring that, at any point in time, the lowest energy state of a given charge operator expectation value is a charge eigenstate. So we work with time-independent states in the Heisenberg picture. 
	
	Let us label the eigenvalues of the charge operator by $n \in \mathbb{N}_0$, and the associated eigenstates as $|n\rangle$, so that
	\be
	\hat{Q} |n\rangle = n |n\rangle \;.
	\ee 
	Note that $|n\rangle$ is usually a family of states as labelled by other distinguishing features, but this is not important for our analysis because we are focusing on the lowest energy representative. Indeed, let us denote by $E_n$ the energy of the lowest energy eigenstate with eigenvalue $n$, so that
	\be
	\langle n|\hat{H}|n\rangle = E_n \;. 
	\ee
	We have normalized the charge eigenstates so that
	\be
	\langle n | n \rangle = 1 \;.
	\ee
	We consider a general state $|\psi\rangle$ which is some arbitrary superposition of the charge eigenstates
	\be
	|\psi\rangle = \sum_n a_n |n\rangle \;.
	\ee
	We denote the energy and charge {\it expectation value} of the state as $E_{\psi}$ and $Q$ respectively
	\be
	\langle \psi|\hat{H}|\psi \rangle = E_{\psi} \;,\;\;\; \langle \psi|\hat{Q}|\psi\rangle = Q \;. 
	\ee
	We restrict to the cases 
	\be
	Q \in \mathbb{N}_0 \;.
	\ee
%	The state is normalized
%	\be
%	\langle \psi | \psi \rangle = \sum_n a^2_n = 1 \;,
%	\label{normpsi}
%	\ee
	
	Let us consider the possible forms of $|\psi \rangle$. We require that it is normalised $\langle \psi | \psi \rangle=1$, and that its charge expectation value is restricted to be $Q$. In terms of the $a_n$, these give the constraints
	\be
	\sum_n a^2_n = 1 \;,\;\;\; \sum_n n \;a_n^2 = Q \;,\;\;\; E_{\psi} = \sum_n E_n a_n^2 \;.
	\label{chaexp}
	\ee
	The simplest solution to (\ref{chaexp}) is that $|\psi\rangle$ is a charge eigenstate, so 	
	\be
	a_Q = 1 \;,\;\;\; a_{n\neq Q} = 0  \;\implies E_{\psi} = E_Q \;.
	\label{cheg}
	\ee
	The next solution we can consider is a two-state superposition, so the only non-vanishing $a_n$ are denoted $a_{n_1}$ and $a_{n_2}$ such that
	\be
	a_{n_1}^2 + a_{n_2}^2 = 1 \;,\;\;\; n_1 a_{n_1}^2 + n_2 a_{n_2}^2 = Q \;,\;\; E_{\psi} = E_{n_1} a_{n_1}^2 + E_{n_2} a_{n_2}^2\;. 
	\ee
	We can use the normalization to eliminate $a_{n_2}$, and then label $a_{n_1}^2=\lambda$, to give
	\be
	Q = n_1 \lambda + \left(1-\lambda\right) n_2 \;, \;\;\; E_{\psi} = E_{n_1} \lambda + \left(1-\lambda\right) E_{n_2} \;.
	\ee
	The normalization gives the constraint
	\be
	0 \leq \lambda \leq 1 \;,
	\label{lamrng}
	\ee
	but otherwise $\lambda$ is unconstrained. Now, requiring that the lowest energy state for a fixed expectation value of the charge operator $Q$ is a charge eigenstate implies
	\be
	E_{n_1 \lambda + \left(1-\lambda\right) n_2} \leq  E_{n_1} \lambda + \left(1-\lambda\right) E_{n_2} \;.
	\label{purcon}
	\ee
	This is equivalent to requiring convexity of the energy spectrum of charge eigenstates. 
	
%	The only thing we should check is that $\lambda$ satisfies (\ref{lamrng}). 
%	We can solve for $\lambda$ in terms of $Q$, $n_1$ and $n_2$ as
%	\be
%	\lambda = \frac{Q - n_2}{n_1 - n_2} \;. 
%	\ee
%	This means that for any $Q$, we can find $n_1$ and $n_2$, subject to the constraints
%	\be
%	n_2 \leq Q \leq n_1 \;,
%	\ee
%	which give an appropriate $\lambda$ satisfying (\ref{lamrng}). 
	
%%%%%%%%%%%%%%%%%%%%%%%%%%%%%%%%%%%%%%%%%%%%%%%%%
\subsection{Convexity in the charge expectation value}
\label{sec:cce2}
%%%%%%%%%%%%%%%%%%%%%%%%%%%%%%%%%%%%%%%%%%%%%%%%%

We can connect these results to the ones in section \ref{sec:ccp}, where fields and actions were considered. Specifically, we show that the energy spectrum of states as a function of the charge expectation value is always convex. This implies convexity in charge eigenvalues if the lowest energy state for a given expectation value is an eigenstate. 

The approach we adopt is to use external current methods. We consider a field $\chi$ which transforms non-linearly under the $U(1)$, so with a shift symmetry. We consider the generating functional with $\chi$ coupling to a current $J_{\chi}$:
\be
	Z[J_{\chi}] = \int {\cal D}\chi\; e^{i\left( S[\chi] + \int \chi \; J_{\chi} \;d^dx \right)} \;,
	\label{Minintj}
\ee
To be clear, in (\ref{Minintj}), by abuse of notation, we are denoting by ${\cal D}\chi$ performing the full path integral over all the fields in the theory. However, we only couple the specific field $\chi$ to a current.
%\footnote{Note that, in general, when specifying an action the formulation may be such that there are multiple fields which all transform under the $U(1)$. For showing convexity, we can choose any of those fields to represent $\chi$. This is because convexity is preserved under linear transformations. However, it is cleanest to define $\chi$ as the particular linear combination which is associated to the $U(1)$. Specifically, the multiple fields will have multiple $U(1)$s associated to them, that are then broken to the specific $U(1)$ through interactions. Then $\chi$ is the field which does not transform under the broken $U(1)$s.} 
Note that, because of the shift symmetry of $\chi$, we could couple it through its derivative $\partial_{\mu}\chi$ to  an external four-current $J_{\chi}^{\mu}$, but it is more convenient to couple $\chi$ itself. 

It is useful to go to Euclidean time and write the generating functional in terms of the Euclidean action $S_E$ as  
\be
	Z[J_{\chi}] = \int {\cal D}\chi \;e^{-S_E[\chi] + \int \chi \; J_{\chi} \;d^{d-1}x d\tau} \;.
	\label{intj}
\ee
We define the connected generating functional
	\be
	W[J_{\chi}] = \log \left( \frac{Z[J_{\chi}]}{Z[0]} \right) \;.
	\ee
In Euclidean time it is manifest that $W[J_\chi]$ is a convex functional of the function $J_\chi$:
	\be
	W[\lambda J^{(1)}_{\chi} + (1-\lambda) J^{(2)}_{\chi} ] \leq \lambda W[J^{(1)}_{\chi}] + (1-\lambda) W[J^{(2)}_{\chi}] \;, \hspace{1em} \text{for $0 \leq \lambda \leq 1$.}
	\ee 
	This follows from Holder's inequality
	\be
	\int f^{\lambda} g^{1-\lambda} d \mu \leq \left(\int f d\mu\right)^{\lambda}\left(\int g d\mu\right)^{1-\lambda} \;,\;\; 0 \leq \lambda \leq 1 \;,
	\label{holdinequ}
	\ee
	which holds for any positive definite measure $d\mu$. We use $f=e^{\int J^{(1)}_{\chi} \chi \;d^{d-1}x d\tau}$ and $g=e^{\int J^{(2)}_{\chi} \chi d^{d-1}x d\tau}$ and note that the measure $d\mu = \frac{1}{Z[0]}e^{-S_E[\chi]}d\chi$ is positive definite. It is important to state that using Holder's inequality here requires integrating over real $\chi$, and in particular real $\dot{\chi}$ in the Euclidean action $S_E[\chi]$. This is subtle due to saddle points being at imaginary $\dot{\chi}$. We discuss this issue in more depth in section \ref{sec:backmink}.

The Euclidean effective action as a function of the classical field $\chi_c$ is defined as the convex conjugate (Legendre transform) of $W[J_{\chi}]$:
	\be
	\Gamma^E_{\mathrm{eff}}[\chi_c] = \sup_{J_\chi}\;\Big[\int \chi_c \; J_{\chi}  \;d^{d-1}x d\tau  - W[J_{\chi}] \Big] \;.
	\label{effacclf}
	\ee
	The convex conjugate of a convex functional is also convex. Therefore, $\Gamma^E_{\mathrm{eff}}[\chi_c]$ is a convex functional of the function $\chi_c$. 

The convexity property of the effective Euclidean action was discovered in~\cite{Iliopoulos:1974ur}, and is textbook material, see for example~\cite{Duncan:2012aja}. It was recently utilised in~\cite{Moser:2021bes}. 	
	
The effective action $\Gamma^E_{\mathrm{eff}}[\chi_c]$ should be treated with care. One point is that, in a Quantum Field Theory with spontaneous symmetry breaking, the effective potential is flat between minima of the potential, but anywhere along this flat direction, apart from the original minima points, cluster decomposition fails (see, for example,~\cite{Duncan:2012aja}). It will also in general be non-local. 

Another subtlety is related to singularities in the action. Such singularities can be understood in two ways. The first is if we consider performing the path integral in the variable $\chi$. In that case, there may be a Jacobian transforming from the fields in the action to $\chi$, that can become singular on certain loci. The simplest example is a weakly-coupled theory where there is a field $\phi$ linearly realising the symmetry $\phi \rightarrow e^{i\xi} \phi$, and we write $\phi = |\phi|e^{i\chi}$. Then the Jacobian develops a logarithmic singularity over the locus $|\phi|=0$. The second way is to perform the integral in the original fields, but then we have that $\chi = -i \left( \log \phi - \log |\phi|\right)$ becomes ill-defined. 

All these subtleties of the effective action can be neglected if we only use it to extract the energy of the minimal energy state of a given charge: Let us restrict to evaluating the effective action on the minimal energy state, the one that breaks spontaneously the \(U(1)\) symmetry $\chi_c = \bar \chi_c $. Our analysis below is valid keeping this state $\bar \chi_c$ completely general. However, we also know its form is 
   \be
   \bar{\chi}_c = m t = -i m \tau \;.
   \label{chstacb}
   \ee
 % There may be an additional constant in the ground state profile, but under the assumption of shift symmetry in $\chi$ this does not affect any energy calculations and can be dropped. 
Since $\chi_c$ (and $\chi$) enjoys a shift symmetry, the effective action is a functional only of its time derivatives. Further, time parity implies that these time derivatives must appear in even powers. This means that the Euclidean action evaluated on the profile $\bar{\chi}_c$ is real. It is also the case that non-localities can be written as higher derivative terms in the action and these vanish on the profile linear in time. Finally, there should be no singularities in the action as long as the state is charged and the theory is interacting. This is because the singularity would be associated to a state allowing an infinite rotation speed, $m \rightarrow \infty$, with finite energy. We do not expect that this can happen in anything other than free theories. We note, however, that we have no proof of this expectation.

Given these simplifications, evaluating the effective action on the charged state  we may write
\be
\Gamma^E_{\text{eff}}[\bar{\chi}_c] = \int {\cal L}_{\mathrm{eff}}^E\left(\dot{\bar{\chi}}_c\right) \;d^{d-1}xd\tau  = {\cal L}_{\mathrm{eff}}^E\left(\dot{\bar{\chi}}_c\right) VT\;,
\label{efftostate}
\ee 
with ${\cal L}_{\mathrm{eff}}^E$ being the Euclidean Lagrangian, that is some real function in $\dot{\bar{\chi}}_c$ (with not necessarily integer powers). In the last step we used that the profile is homogenous, and denote $\int d^{d-1}x d\tau = VT$
The effective Euclidean Lagrangian is convex in $\bar{\chi}_c$ and therefore, since the derivative is a linear operation, also in $\dot{\bar{\chi}}_c$.

%%%%%%%%%%%%%%%%%%%%%%%%%%%%%%%%%%%%%%%%%%%%%%%%%%%%%%%%%%%%%%%
\subsubsection*{Rotating back to Minkowski time}
\label{sec:backmink}
%%%%%%%%%%%%%%%%%%%%%%%%%%%%%%%%%%%%%%%%%%%%%%%%%%%%%%%%%%%%%%%

We now would like to rotate back to Minkowski time. To track convexity of the Lagrangian through the time rotation it is convenient, though not required in general, to assume it is differentiable in $\dot{\bar{\chi}}_c$. We can then write
\be
\frac{\partial^2}{\partial\dot{\bar{\chi}}_c^2} {\cal L}_{\mathrm{eff}}^E\left(\left( \dot{\bar{\chi}}_c\right)^2 \right) \geq 0 \;.
\label{convexchidot}
\ee
We would now like to translate this to a statement about convexity of the Minkowski Lagrangian in $m$. This can be done through a series of variable changes.\footnote{One could simply switch $\partial_{\dot{\bar{\chi}}_c} \rightarrow i \partial_m$, but to be clearer we perform the variable changes such that the variable domains and the Lagrangian are manifestly always real, where convexity is well-defined.} First we switch to variable $y \equiv \left( \dot{\bar{\chi}}_c\right)^2$. We can then write~(\ref{convexchidot}) as
\be
\left(\frac{\partial}{\partial y} + 2 y \frac{\partial^2}{\partial y^2} \right) {\cal L}_{\mathrm{eff}}^E\left(y \right) \geq 0 \;.
\label{convexy}
\ee
We now switch to Minkowski time. This means we switch to the variable $z = -y$, and switch to the Minkowski Lagrangian using~(\ref{eutomink}) (the latter just corresponds to an overall minus sign in the Lagrangian):
\be
\left(\frac{\partial}{\partial z} + 2 z \frac{\partial^2}{\partial z^2} \right) {\cal L}_{\mathrm{eff}}\left(z \right) \geq 0 \;.
\label{convexz}
\ee
Finally we switch to the variable $m = \sqrt{z}$, which gives
\be
\frac{\partial^2}{\partial m^2} {\cal L}_{\mathrm{eff}}\left(m\right) \geq 0 \;.
\ee
Therefore, ${\cal L}_{\mathrm{eff}}$ is convex in $m$. Note that all the convexity statements are over a certain domain in each variable. This is left implicit in the analysis above. 

The rotation back to Minkowski time ensures that the final convexity result is for a real-valued Lagrangian in the real-valued parameter of $\dot{\bar{\chi}}_c = \partial_{t}\bar{\chi}_c = m$. This is the final important result for us, from which we will proceed. 

However, the path to reach this result, through the Euclidean time calculation, has some important subtleties. The Euclidean path integral (\ref{intj}) was performed over a real $\chi$ with a real Euclidean time derivative $\partial_{\tau}\chi$. This is required to ensure that the integral is convergent, with a positive Euclidean action $S_{E}[\chi]>0$, and is the standard analytic continuation method. It is also required in order to use Holder's inequality (\ref{holdinequ}). However, the ground state ansatz (\ref{chstacb}) has $\partial_{\tau}\overline{\chi}_c$ imaginary, not real. In the path integral this means that, at least for a sufficiently weakly-coupled theory, there are saddle points for which $\partial_{\tau}\chi$ is imaginary. Our assumption is that such imaginary saddles are still captured by the integral over the real line. The point is that the rotation to Euclidean time demands that $\chi$ is also analytically continued and one must now integrate over some path in the complex plane. The integration path we take, which is the only one that can be manifestly controlled, is the real line. This path can, assuming no pole obstructions, be deformed to one which goes through the saddle, while still maintaining reality of the action. Such a deformed path, going through the saddle, is known as a Lefschetz Thimble, see \cite{Scorzato:2015qts} for a review. This whole subtlety is universal to path integrals in Euclidean time, and is a practical problem for Lattice studies. We have nothing new to add on this matter, and only assume (as standard) that after the final rotation back to Minkowski time, the resulting effective Lagrangian captures all the physics correctly. 

We should emphasise that it is really a central, crucial step in our analysis to evaluate the Euclidean path integral over the real line for $\chi$ and $\dot{\chi}=\partial_{\tau}\chi$. This is the crucial difference from attempting more direct approaches to showing convexity. For example, it is simple to show convexity of the partition function in the chemical potential ($m$), but this only implies convexity of the energy in charge in a thermodynamic limit. Another approach one could take is to evaluate the Hamiltonian directly over a fixed charge state, similar to the approach of \cite{Badel:2019oxl}, but this has a related issue of a factor of $i$ which obstructs using Holder's inequality. We discuss these approaches in more detail in appendix \ref{sec:simpro}. It may be that these approaches could be used to generalise our results through a better understanding of the appropriate path to take in the complex $\chi$ plane.   

%%%%%%%%%%%%%%%%%%%%%%%%%%%%%%%%%%%%%%%%%%%%%%%%%%%%%%%%%%%%%%%
\subsubsection*{Convexity in charge}
%%%%%%%%%%%%%%%%%%%%%%%%%%%%%%%%%%%%%%%%%%%%%%%%%%%%%%%%%%%%%%%

   We stay now in Minkowski time. By definition, the charge \(Q\) under the \(U(1)\) is the conjugate momentum to $\dot{\bar{\chi}}_c=m$, so a Legendre transform will give us the Hamiltonian
	\be
	{\mathrm H}_{\mathrm{eff}}(Q) = Q\; m - V {\cal L}_{\mathrm{eff}}(m) \;,
	\label{legtoH}
	\ee
	where the charge is defined as
	\be
	 Q \equiv V \frac{\partial{\cal L}_{\mathrm{eff}}}{\partial m} \;.
	 \label{qdeflm}
	\ee
	Since a Legendre transform preserves convexity, ${\mathrm H}_{\mathrm{eff}}(Q)$ is a convex function of $Q$.
  
 We emphasise that the results of this section only show convexity in the charge expectation value, so $Q = \left< Q \right>$, not for the charge eigenvalues. The expectation value is the only quantity that is probed through an external current approach, and is the best that one can obtain using such an approach. Convexity in the expectation value implies convexity in the eigenvalues if the lowest energy state is an eigenstate. Therefore, we have arrived at the same result as section \ref{sec:conpure}, but this time through actions, fields and external currents.

%%%%%%%%%%%%%%%%%%%%%%%%%%%%%%%%%%%%%%%%%%%%%%%
\section{Discussion}
\label{sec:dis}
%%%%%%%%%%%%%%%%%%%%%%%%%%%%%%%%%%%%%%%%%%%%%%%

We studied the spectrum of charged operators in \textsc{cft}s with a $U(1)$ global symmetry. We defined the large $f$ regime: This is a regime in parameter space, and in the Hilbert space, so for example for certain ranges of charges, where the charged states dual to the charged operators have a Goldstone boson excitation about them. That is, a field $\pi$ which realises the $U(1)$ symmetry non-linearly. More precisely, we do not even require a field but some modes without a gap, just the same as shown in Goldstone's theorem. The parameter $f$ is then identified with the symmetry breaking scale associated to the Goldstone boson, and we require it to be large relative to the sphere radius of the \textsc{cft} (otherwise, one could just say there is no Goldstone boson). This identifies a new regime in \textsc{cft}s which can be studied generally, and which does not coincide with the large charge regime as introduced in \cite{Hellerman:2015nra}.  

We showed that in the large $f$ regime, there is a convex spectrum of states as a function of charge, and therefore a convex spectrum of operators. The  result is then that: if there is a Goldstone boson, there is a convex spectrum.\footnote{In fact, we find not only convexity, but a certain lower bound on the amount of convexity (\ref{sublum}).} More precisely, what we find are (possibly multiple) convex families of states/operators, where such a family is defined by the embedding of the Goldstone boson into the degrees of freedom which describe the \textsc{cft} in the large $f$ regime of interest (there may be multiple such regimes with different degrees of freedom). So a family is the set of lowest dimension operators, or lowest energy states, which have the same Goldstone boson associated to their dual states. For example, in a weakly-coupled theory with a complex field $\phi$, $\phi^n$ with $n$ a (large relative to one) integer, is a family where the Goldstone boson is the same, being given by the phase of $\phi$. If $\phi$ is also the field with the largest $U(1)$ charge in the theory, that family would be lowest energy, and therefore must be convex.

We note though that, although we believe our results are correct, there are some subtleties which could provide loopholes and ways out. We note them here. One is that we assumed, with justifications, in section \ref{sec:gbl} that $\gamma$ cannot be negative and small. Also, in the case when there is no gap between the Goldstone modes and other fields, studied in section \ref{sec:gbl}, we wrote a theory where the Goldstone couples to these fields in a general way (at leading order in $f$). We gave arguments, but did not prove, that this is the most general way to couple the Goldstone to other sectors. Related to these is the issue that we do not have a general formulation of the key objects in the analysis, the Goldstone boson and the scale $f$, in terms of general properties of \textsc{cft}s such as currents and OPE's. This means that it is not simple to define when a \textsc{cft} is in the large $f$ regime, although there should always exist one as long as the $U(1)$ symmetry is broken. 

The results do not imply precisely the Charge Convexity Conjecture of \cite{Aharony:2021mpc}. There are three main reasons: The first is that it is not guaranteed that there exists a large $f$ regime at charges which are not parametrically large in any parameter of the $\textsc{cft}$. However, it is natural to expect, and this is true in all the examples we know, that such a regime should appear already at $Q \gg 1$, rather than being controlled by some other parameters.\footnote{There are also some exceptional cases, such as the theory of free fermions, where there is no symmetry breaking and no Goldstone boson at all. See, for example \cite{Komargodski:2021zzy} for an analysis.} The second reason is that it is not necessarily the case that the lowest dimension operators within a charge range share the same Goldstone boson realisation. So, the dual states to the operators would break the symmetry differently: they would have different combinations of fields obtaining expectation values. Convexity follows only when the lowest dimension operators of different charges all break the symmetry in the same way, so share a Goldstone boson. It is possible to construct models where this happens only at parametrically large charge, as shown in \cite{Sharon:2023drx}. 
The third reason is that we studied only homogenous states, and so scalar operators. We have nothing to say in this work about non-scalar operators, though our approach can still be used to study them.\footnote{Note that we also restricted to operators which do not have a moduli space, but if they do then they are BPS which implies convexity by the BPS bound.}   

In terms of the Weak Gravity Conjecture \cite{Arkani-Hamed:2006emk}, or more precisely the Positive Binding Conjecture \cite{Aharony:2021mpc}, our results present strong evidence for certain aspects of them, but leave some open questions. The evidence is for the statement that the particle with the largest charge-to-mass ratio must have positive self-binding. This is motivated by thinking of multi-particle states as $\phi^n$ type families, which share a Goldstone embedding. So the requirement we need for convexity is only that this family are of lowest dimension for their charge. The counter example to charge convexity in \textsc{cft}s of \cite{Sharon:2023drx}, is analogous to this particle having parametrically large charge. It is, of course, not clear though that this could happen in a holographic context with Einstein gravity. The main open question for a holographic interpretation of our results is the dual of the Goldstone boson. 

We showed that, in the large $f$ regime, non-convexity is mapped to an inconsistency. In the case when there is a gap between the Goldstone momentum modes and the next heaviest states, this inconsistency corresponds to wrong sign kinetic terms for the Goldstone. More generally, including when there is no gap, the inconsistency is an instability of the state, even though it is the lowest energy states in a charge superselection sector and so cannot decay. It would be interesting to find other inconsistencies or ways towards convexity. We discuss some such ideas in appendix \ref{sec:UC}. In particular, we showed in section \ref{sec:cce}, that convexity can be mapped completely generally to the statement that the lowest energy state of a given charge operator {\it expectation value} should be a charge eigenstate.

%It is informative to see how the general result of this paper matches onto data from specific theories. 
%	In~\cite{Palti:2022unw} it was shown that in two dimensions it is possible to parametrically delay convexity even for the scalar operator sector. This matches the fact that in two dimensions there are no Goldstone bosons. It is also known that \textsc{cft}s which are not unitary and unstable, such as the $O(N)$ model in five dimensions, can violate convexity of charged operators~\cite{Aharony:2021mpc,Moser:2021bes}. This is matched onto the instabilities we find in our analysis. 
		
	It is natural to extended our analysis to the case of non-Abelian symmetries, and to non-scalar operators.
   These two generalizations are related, since we know that for a general representation the lowest state is not homogeneous~\cite{Alvarez-Gaume:2016vff,Hellerman:2017efx,Hellerman:2018sjf,Banerjee:2019jpw}.
  Also, in \cite{Palti:2022unw} the study of convexity in the case of multiple $U(1)$ symmetries was initiated. It was proposed that convexity can be quantified in terms of a lattice index. It would be interesting to extend the results of this paper to multiple $U(1)$s and the associated index.

Our analysis is independent of the holographic AdS radius, and so holds arbitrarily close (in terms of curvature scales) to the flat space limit of AdS. However, it was argued in~\cite{Lust:2019zwm}, that the flat space limit should be thought of as being at infinite distance from any AdS space, and in this sense it is still not clear if one can extrapolate any statement in AdS to flat space. If we assume that flat space can be sufficiently well approximated by a sufficiently weakly-curved AdS, then our results may be relevant also for the Positive Binding Conjecture in flat space.\footnote{More precisely, the flat space limit of positive binding was shown in \cite{Andriolo:2022hax} to match, as expected, the repulsive force conjecture \cite{Palti:2017elp,Heidenreich:2019zkl}.}

	\vskip 20pt
	\noindent {\bf Acknowledgements:} We deeply thank Ofer Aharony, Gabriel Cuomo, Zohar Komargodski, Joao Penedones, Adar Sharon, and Masataka Watanabe for very useful discussions and explanations. The work of EP was supported by the Israel Science Foundation (grant No. 741/20) and by the German Research Foundation through a German-Israeli Project Cooperation (DIP) grant "Holography and the Swampland". We thank SwissMAP and the Swiss National Science Foundation for hospitality during the workshop "Large Charge in Les Diablerets", where this work was initiated.

\appendix

%%%%%%%%%%%%%%%%%%%%%%%%%%%%%%%%%%%%%%%%%%%%%%%
	\section{Alternative paths to convexity}
	\label{sec:simpro}
%%%%%%%%%%%%%%%%%%%%%%%%%%%%%%%%%%%%%%%%%%%%%%%

In this appendix we discuss some alternative paths towards showing convexity of the spectrum of charge states (and therefore operators). To the best of our understanding, these paths are obstructed, but they have useful intermediate results, and are informative. 

The first path is through first showing that the partition function is convex in the chemical potential~\cite{Israel:1979}.
We consider a CFT, assuming only a $U(1)$ global symmetry, and the existence of a Hamiltonian. For an inverse temperature $\beta$, the finite temperature partition function is
	\be
	Z = \mathrm{Tr} \left[e^{-\beta \hat{H}} \right] \;.
	\ee
We can now consider this in a chemical potential $m$.\footnote{Note that instead of a chemical potential we could equally well couple the Noether current of the charge to a background gauge field. The time component of the background gauge field then acts effectively as a chemical potential.} This gives
	\be
	Z[m] = \mathrm{Tr} \left[e^{-\beta \left(\hat{H} - m \hat{Q} \right) } \right] \;.
	\ee
	Let us define $P$ as the negative of the grand potential $\Omega$, so $P=-\Omega$, and so
\be
 P[m] \equiv \mathrm{log\;}Z[m] \;.
 \ee
Convexity of $P[m]$ then follows from Holder's inequality. Specifically, we write $m = \lambda m_1 + \left(1-\lambda\right) m_2$ and in (\ref{holdinequ}) take 
\be
f = e^{\beta m_1 \hat{Q}} \;,\;\; g = e^{\beta m_2 \hat{Q}} \;,\;\; d\mu = e^{-\beta \hat{H}}\;.
\ee
This yields,
\be
P[\lambda m_1 + \left(1-\lambda\right) m_2] \leq \lambda P[m_1] + \left(1-\lambda \right)P[m_2]\;.
\ee
In a thermodynamic setting, the (Helmholtz free) energy $F$, as a function of the charge $Q$, is the Legendre transform of $P$, exchanging the chemical potential for the dual charge. Explicitly,
\be
F = m Q - P \;,\;\; Q = \frac{\partial P}{\partial m}\;.
\ee
Since Legendre transforms preserve convexity, this yields convexity of the energy as a function of charge. This very simple and general analysis shows that convexity in charge of the operator spectrum of a CFT should hold in any regime which is behaving thermodynamically. For example, the large-charge regime. 

Away from a thermodynamic limit, the relation between the grand potential and energy is a Fourier transform rather than a Legendre transform. To see this we note that we can extract the energy $E_{Q_0}$, of the lowest energy state of a given charge $Q_0$, by inserting a delta function into the partition function
\be
\lim_{\beta \rightarrow \infty} e^{-\beta E_{Q_0}} = \lim_{\beta \rightarrow \infty} \mathrm{Tr} \left[\delta\left(\hat{Q} - Q_0\right)e^{-\beta \hat{H}} \right] = \lim_{\beta \rightarrow \infty} \int d\theta \;\mathrm{Tr} \left[e^{i\left(\hat{Q} - Q_0\right)\theta}e^{-\beta \hat{H}} \right] \;.
\label{Hdelta}
\ee
We could now attempt to use Holder's inequality to extract convexity in $Q_0$, but we see that it appears with a factor of $i$ in the exponent. Taking the absolute value in (\ref{holdinequ}) would then remove the charge dependence. 

It is interesting to try and repeat the analysis by using a real representation of the delta function, for example
\be
\delta\left(\hat{Q} -Q_0\right) = \lim_{b \rightarrow 0} \frac{1}{|b|\sqrt{\pi}}e^{-\frac{\left(\hat{Q} -Q_0 \right)^2}{b^2}} \;.
\ee
In that case one finds that the energy as a function of charge behaves as
\be
E_{Q_0} \sim \lim_{\beta \rightarrow \infty} \lim_{b \rightarrow 0} \frac{1}{\beta}\left[ \frac{Q_0^2}{b^2} - P\left(Q_0,b\right) + \log |b| \right] \;,
\ee
where, by Holder's inequality, $P\left(Q_0,b\right)$ is a convex function of $Q_0$. While $Q_0^2$ is convex in $Q_0$, the difference between two convex functions is not necessarily convex, and so we cannot deduce convexity from this approach either without some further input.

The second approach is essentially a Lagrangian dual formulation of the same idea. Instead of evaluating the partition function, we could evaluate the Hamiltonian directly on a charge eigenstate $|Q_0\rangle$ of charge $Q_0$, so
\be
\langle Q_0 | e^{-\beta \hat{H}} | Q_0 \rangle = \int_{| Q_0 \rangle}^{| Q_0 \rangle} e^{-\beta S_E} \;,
\ee
where we denoted by the integral limits the value of the fields in the state $| Q_0 \rangle$. As shown in \cite{Badel:2019oxl}, taking a field $\chi$ transforming non-linearly under the $U(1)$, the path integral can be written as
\be
\int e^{-\beta \left(S_E + i \int \rho_0 \dot{\chi}  \;d^{d-1}x d\tau \right)} \;,
\ee
where $\rho_0$ is the charge density associated to $Q_0$. Again, as in (\ref{Hdelta}), if we tried to use Holder's inequality to show convexity in $\rho_0$, the factor of $i$ would remove the charge dependence. Note that, as in our analysis in the main text, writing $\dot{\chi} = -im$, from (\ref{chstacb}), would remove the factor of $i$. However, the Euclidean path integral requires $\dot{\chi}$ real to be controlled though the action, and so we must integrate over that path. 

%%%%%%%%%%%%%%%%%%%%%%%%%%%%%%%%%%%%%%%%%%%%%%%%%
\subsection{Unitarity, causality and convexity}
\label{sec:UC}
%%%%%%%%%%%%%%%%%%%%%%%%%%%%%%%%%%%%%%%%%%%%%%%%%

We have seen that convexity can be related to the positivity of kinetic terms. We can think of this in terms of the Goldstone boson $\pi$, or more directly in terms of the kinetic terms for $\chi$. There are known constraints on the positivity of kinetic terms for shift symmetric fields \cite{Adams:2006sv}. It is therefore interesting to consider whether the two can be related. We do find some relation, though we do not know how general it is.\footnote{We note here that also in \cite{Hellerman:2017veg} a relation between the correction to the dimension of operators at large charge, and the positivity coming from scattering amplitudes was pointed out.}

In section \ref{sec:to2m} we discussed an example class of weakly-coupled theories. In this case, the Lagrangian was quartic in $m$. Since $m$ arises from the derivative of $\chi$, this means that the effective Lagrangian for $\chi$, after integrating out the $f$ field, is quartic in the derivatives of $\chi$. 

Let us consider a completely general Lagrangian that is quartic in $m$, and see what are the constraints on convexity. We write
\be
{\cal L}\left(m\right) = a + b m^2 + c m^4 \;.
\ee
Now we should demand positivity of charge, $\frac{\partial {\cal L}}{\partial m}>0$, which yields
\be
4 c m^2 + 2b > 0 \;.
\label{psch}
\ee 
Then evaluating the second derivative gives
\be
\frac{\partial^2{\cal L}}{\partial m^2}  = 2b + 12 c m^2 > 8 c m^2 \;,
\ee
where we used (\ref{psch}). We therefore see that convexity is mapped onto the condition 
\be
c > 0 \;,
\label{cuc}
\ee
where $c$ appears in the Lagrangian as the coefficient of the quartic derivative term for $\chi$, so
\be
{\cal L}\left(\chi\right) \supset c\left(\partial \chi \right)^4 \;.
\ee
In \cite{Adams:2006sv} it was shown that for a shift-symmetric scalar, as is $\chi$, unitarity and causality imply precisely the condition (\ref{cuc}) in the coefficient of the four-derivative term. 

It is not clear what to make of this connection. There are some crucial differences from the analysis of \cite{Adams:2006sv}. First, we are on a curved background, rather than flat space. Second, we are in a regime where the fourth-derivative term is not subdominant to the second-derivative term. Nonetheless, the connection is striking. 

Perhaps even more generally, we could consider causality constraints on the Goldstone boson $\pi$. These would need to be in the Lorentz symmetry background. An analysis of such constraints was performed recently in \cite{Creminelli:2022onn}. It would be very interesting to see if this could be applied. 

\section{Stability with light radial mode}
\label{sec:two-field-stability}
%%%%%%%%%%%%%%%%%%%%%%%%%%%%%%%%%%%%%%%%%%%%%%%%%%%%%%%%%%

We want to study the stability condition for a system in which a light radial mode is present together with the Goldstone.
This is for example the case on the cylinder when the expectation value of the radial mode is comparable with the scale fixed by the radius of the sphere that controls the amplitude of the excitations over the ground state.

It is convenient to start from a system in which the \(U(1)\) symmetry is realized linearly as acting on a complex field \(\phi = a e^{i \chi}\).
Generically, the action will be a function of the Lorenz-invariant combination \(\del_{\mu} \phi^{*} \del^{\mu} \phi\) and of the absolute value \(\phi^{*} \phi\):
\begin{equation}
  L^{(2)} = L^{(2)}(\del_{\mu} \phi^{*} \del^{\mu} \phi, \phi^{*} \phi ) .
\end{equation}
The fixed-charge ground state realizing the breaking of the \(U(1)\) takes the helical form
\begin{equation}
  \phi(t,x) = A e^{i m t} .
\end{equation}
The constants \(A\) and \(m\) are determined by the \textsc{eom}, which read
\begin{equation}
  \begin{dcases}
    \frac{d L^{(2)}(A^2 m^2, A^2)}{d A} = 0 \\
    \frac{d L^{(2)}(A^2 m^2, A^2)}{d m} = \rho ,\\
  \end{dcases}
\end{equation}
where \(\rho\) is the charge density.
The energy of the ground state is given by the usual Legendre transform:
\begin{equation}
  E(Q) = Q m - \mathrm{Vol} L^{(2)}(A^2 m^2, A^2),
\end{equation}
and
\begin{equation}
  \frac{d E(Q)}{dQ} = m
\end{equation}
so that the second derivative of the energy with respect to the charge is
\begin{equation}
  \frac{d^2 E(Q)}{dQ^2} = \frac{d m}{dQ} .
\end{equation}

We can think of \(m\) as a chemical potential and use it as our control variable.
Then the equation of motion for \(A\) is solved by a function \(A = A(m)\) and, substituting into the expression of \(L^{(2)}\), we can define a function of \(m\) alone:
\begin{equation}
  L(m) = \left. L^{(2)}(A(m)^2 m^2, A(m)^2 ) \right|_{\frac{d L^{(2)}(A^2 m^2, A^2)}{d A} = 0} ,
\end{equation}
and the energy as its Legendre transform
\begin{equation}
  E(Q) = Q m - V L(m) .
\end{equation}
so that
\begin{align}
  \frac{d E(Q)}{dQ} = m  \;,\;\;\frac{d^2 E(Q)}{dQ^2} = \frac{d m}{dQ} = \left( \frac{d Q}{dm}  \right)^{-1} = \left( \frac{d^2 L}{d m^2}  \right)^{-1} .
\end{align}
This would be the starting point for an analysis in which the radial mode is integrated out.
However we are assuming here that the fluctuations of the field \(a\) over the \textsc{vev} \(A\) are not suppressed so, in order to avoid the proliferation of higher-derivative terms, we will keep both modes and use the Lagrangian as function of two fields.

Not integrating out \(a\) allows us to assume that the Lagrangian \(L^{(2)}\) can be written in the standard form
\begin{equation}
   L^{(2)}(\del_{\mu} \phi^{*} \del^{\mu} \phi, \phi^{*} \phi )  =  \del_{\mu} \phi^{*} \del^{\mu} \phi - V( \phi^{*} \phi )
\end{equation}
and the equations of motion take the form
\begin{equation}
  \begin{dcases}
    m^2 - V'(A^2) = 0 \\
    2 A m = \frac{Q}{V} .
  \end{dcases}
\end{equation}
Deriving both equations with respect to \(m\) we find an expression for the derivative \(A'(m)\) and \(Q' (m)\):
\begin{equation}
  \begin{dcases}
    \frac{dA}{dm} = \frac{Q}{2 A^3 V''(A^2)} \\
    \frac{dQ}{dm} = \frac{Q}{m}\left( 1+ \frac{Q m}{A^4 V''(A^2)}  \right)
  \end{dcases}
\end{equation}
and we can write the second variation of the energy with respect to the charge as
\begin{equation}
  \frac{d^2 E(Q)}{dQ^2} = \frac{d m}{dQ} = \left(  \frac{dQ}{dm} \right)^{-1} = \frac{m}{Q}\left( 1+ \frac{Q m}{A^4 V''(A^2)}  \right)^{-1}.
\end{equation}
Note that this can be equivalently expressed in terms of \(L(m)\) as
\begin{equation}
  \frac{d^2L}{d m^2} =  \frac{Q}{m}\left( 1+ \frac{Q m}{A^4 V''(A^2)}  \right).
\end{equation}

We want to relate the sign of this quantity to the stability of the fluctuations over the ground state.
To do this, we write the field as
\begin{equation}
  \phi = e^{i m t} (A + \hat \varphi)  
\end{equation}
and expand at second order in \(\hat \varphi\).
The inverse propagator for \(\hat \varphi \) in Fourier space is
\begin{equation}
  \Delta^{-1} = \frac{Q}{m A^2}
  \begin{pmatrix}
    -\frac{2 A^4 m V''(A^2)}{Q}  + \frac{1}{2}  (\omega^2 - k^2) & -  i m  \omega \\
     i m  \omega &  \frac{1}{2}  (\omega^2 - k^2)
  \end{pmatrix},
\end{equation}
where \(k^2\) stands for the eigenvalues of the Laplacian on the spatial part.
The dispersion relations for the two modes are obtained from the condition \(\det(\Delta^{-1}) = 0\):
\begin{equation}
  \det(\Delta^{-1}) \propto \omega^4 - \left(  2 k^2 + 4 m^2 + 4 \frac{A^4 m}{Q} V''(A^2)  \right) Q \omega^2 + k^2 \left(  k^2 + \frac{4 A^4 m}{Q} V''(A^2)  \right)  = 0
\end{equation}
For \(k = 0\) the equation reduces to
\begin{equation}
  \omega^4 - \left( 4 m^2 + 4 \frac{A^4 m}{Q} V''(A^2)  \right)   Q \omega^2  = 0
\end{equation}
which admits the solutions
\begin{align}
  \omega &= 0  & \omega^2 &=  4 m^2 Q + 4 A^4 m V''(A^2)
\end{align}
which correspond to the Goldstone and the massive mode respectively.

Stability of the solution requires  the frequencies \(\omega\) to be real for any value of \(k\) or, equivalently, the equation for \(\omega^2\) to admit two real positive solutions.
Using the fact that a quadratic equation of the form \((\omega^2)^2 - b \omega^2 + c = 0\) admits two real positive solutions if and only if \(b >0\) and \(b^2 \ge 4 c > 0\), we find that the stability of the fluctuations requires
\begin{equation}
  \begin{dcases}
    2 k^2 + 4 m^2 + 4 \frac{A^4 m}{Q} V''(A^2) > 0 \\
    k^2 + \frac{4 A^2 m}{Q} V''(A^2) > 0 \\
    Q^2( k^2 + m^2 ) + 2 A^4 Q m V''(A^2) + A^8 (V''(A^2))^2 \ge 0
  \end{dcases}
\end{equation}
We have seen that the second derivative of the effective action with respect to \(m\) (or, equivalently the second derivative of the energy with respect to the charge) is related to \(V''(A^2)\) by
\begin{equation}
  \frac{d^2L}{d m^2} =  \frac{Q}{m}\left( 1+ \frac{Q m}{A^4 V''(A^2)}  \right),
\end{equation}
so we can rewrite the stability equations as conditions for \(d^2 L / dm^2\) and after some algebra we find
\begin{equation}
  \frac{d^2L}{d m^2} \ge \frac{Q}{m} > 0
\end{equation}
which is precisely the same condition to obtain a stable and causal Goldstone description in section \ref{sec:gbl}.

	\bibliographystyle{JHEP}
	\bibliography{susyswamp}
	%$\bibliographystyle{custom1}
\end{document}